\documentclass[aps,pra,10pt,twocolumn,amsmath,amssymb,showpacs,
longbibliography,superscriptaddress]{revtex4-1}

\usepackage{amsmath,amssymb,amsfonts} 
\usepackage{graphicx}

\usepackage{hhline}
\usepackage[latin1]{inputenc}
\usepackage{subfigure}
\usepackage[colorlinks=true,citecolor=blue,linkcolor=blue]{hyperref}
\usepackage{bm}
\usepackage{bbm}
\usepackage{nicefrac}
\usepackage{wasysym}

\newcommand{\eq}[1]{Eq.~(\ref{eq:#1})}
\newcommand{\eqs}[2]{Eqs.~(\ref{eq:#1}) and~(\ref{eq:#2})}
\newcommand{\equ}[1]{Equation~(\ref{eq:#1})}

\newcommand{\bk}{{\bm k}}

\newcommand{\bn}{{\bm n}}

\newcommand{\bO}{{\bm \Omega}}
\newcommand{\bnab}{{\bm \nabla}}

\def\nn{\nonumber\\}

\def\beq{\begin{equation}}
\def\eeq{\end{equation}}
\def\bea{\begin{eqnarray}}
\def\eea{\end{eqnarray}}

\def\ket#1{\vert#1\rangle}

\def\me#1#2#3{\langle#1\vert#2\vert#3\rangle}
\def\wt#1{\widetilde{#1}}

\def\T{\mathcal T}
\def\P{\mathcal P}

\def\kk{{\bm k}}
\def\KK{{\bm K}}
\def\kz{K_z}
\def\qq{{\bm q}}
\def\rr{{\bm r}}

\def\GG{{\bm G}}

\def\Z2{\mathbbm{Z}_2}

\def\la{\langle\kern-2.0pt\langle}
\def\ra{\rangle\kern-2.0pt\rangle}

\def\zhat{\hat{\bf z}}

\begin{document}

\title{ Composite Weyl nodes stabilized by screw symmetry with and
  without time reversal invariance}

\author{Stepan S. Tsirkin} \affiliation{Centro de F{\'i}sica de
  Materiales, Universidad del Pa{\'i}s Vasco, 20018 San Sebasti{\'a}n,
  Spain} 

\author{Ivo Souza} \affiliation{Centro de F{\'i}sica de Materiales,
  Universidad del Pa{\'i}s Vasco, 20018 San Sebasti{\'a}n, Spain}
\affiliation{Ikerbasque Foundation, 48013 Bilbao, Spain}

\author{David Vanderbilt} \affiliation{Department of Physics and
  Astronomy, Rutgers University, Piscataway, New Jersey 08854-8019,
  USA}

\date{\today}
\begin{abstract}
  We classify the band degeneracies in three-dimensional crystals with
  screw symmetry $n_m$ and broken $\P*\T$ symmetry, where $\P$ stands
  for spatial inversion and $\T$ for time reversal.  The generic
  degeneracies along symmetry lines are Weyl nodes: Chiral contact
  points between pairs of bands. They can be single nodes with a
  chiral charge of magnitude $\vert\chi\vert=1$ or composite nodes
  with $\vert\chi\vert=2$ or $3$, and the possible $\chi$ values only
  depend on the order $n$ of the axis, not on the pitch $m/n$ of the
  screw.  Double Weyl nodes require $n=4$ or 6, and triple nodes
  require $n=6$.  In all cases the bands split linearly along the
  axis, and for composite nodes the splitting is quadratic on the
  orthogonal plane. This is true for triple as well as double nodes,
  due to the presence in the effective two-band Hamiltonian of a
  nonchiral quadratic term that masks the chiral cubic dispersion.  If
  $\T$ symmetry is present and $\P$ is broken there may exist on some
  symmetry lines Weyl nodes pinned to $\T$-invariant momenta, which in
  some cases are unavoidable. In the absence of other symmetries their
  classification depends on $n$, $m$, and the type of $\T$ symmetry.
  With spinless $\T$ such $\T$-invariant Weyl nodes are always double
  nodes, while with spinful $\T$ they can be single or triple nodes.
  $\T$-invariant triples nodes can occur not only on 6-fold axes but
  also on 3-fold ones, and their in-plane band splitting is cubic, not
  quadratic as in the case of generic triple nodes.  These rules are
  illustrated by means of first-principles calculations for hcp
  cobalt, a $\T$-broken, $\P$-invariant crystal with $6_3$ symmetry,
  and for trigonal tellurium and hexagonal NbSi$_2$, which are
  $\T$-invariant, $\P$-broken crystals with 3-fold and 6-fold screw
  symmetry respectively.
\end{abstract}
\maketitle

\section{Introduction}

The study of degeneracies in the energy spectrum of crystals has a
long history in the band theory of solids. The early works focused on
the consequences of
symmetry~\cite{bouckaert-pr36,herring-pr37a,herring-pr37b}, and it was
only much later that the topological aspects of the problem began to
be appreciated~\cite{nielsen-npb81}.  The interplay between topology
and crystal symmetry can be particularly interesting.  For example,
Michel and Zak~\cite{michel-prb99} used an argument based on the
periodicity of reciprocal space to show that nonsymmorphic symmetries
(screw axes and glide planes) necessarily lead to degeneracies on
symmetry lines and planes in the Brillouin zone (BZ).

In recent years the study of band crossings has been reinvigorated by
the discovery of gapless topological phases such as Weyl and Dirac
semimetals, where the presence of degeneracies near the Fermi level
can lead to striking observable
effects~\cite{turner-book13,armitage-arxiv17}.  A knowledge of the
symmetry conditions under which certain types of degeneracies become
possible, or even unavoidable, can greatly simplify the search and
analysis of candidate materials.

Our focus here is on Weyl nodes, i.e., isolated twofold degeneracies
that occur in three-dimensional (3D) band structures without having to
fine-tune the Hamiltonian.  In the simplest and most common case, the
two bands split linearly in all directions away from the
node~\cite{herring-pr37b}.  Such contact points are the generic
degeneracies in bulk crystals with broken $\P*\T$ symmetry, where $\P$
and $\T$ denote spatial inversion and time-reversal symmetry
respectively. (If the combined $\P*\T$ symmetry is present, the bands
are Kramers-degenerate everywhere in the BZ, and additional isolated
degeneracies are known as Dirac nodes.)  Weyl nodes are chiral, acting
as monopole sources and sinks of Berry curvature in the BZ, and when
the quantized Berry flux through some of the Fermi-surface sheets is
nonzero the material is classified as a Weyl
(semi)metal~\cite{turner-book13}.

Weyl nodes are topologically protected by the discrete translational
symmetry of the lattice (they can only be gapped by anihilating with
other Weyl nodes of opposite chirality), and no further symmetries are
needed for their existence.  Nevertheless, the presence of other
symmetries affects their location and characteristics.  For example,
4-fold symmetry can stabilize Weyl nodes along a symmetry axis in the
BZ, and in some cases the bands split quadratically in the directions
perpendicular to the axis (but still linearly along the
axis)~\cite{herring-pr37b,heikkila-jetp10,fang-prl12}.  Such quadratic
touchings may be regarded as consisting of two linear Weyl nodes of
the same chirality brought together by rotational
symmetry~\cite{heikkila-jetp10,fang-prl12}. Their chiral charge is
$\chi=\pm 2$, and for that reason they have become known as ``double
Weyl nodes''~\cite{fang-prl12}.  Furthermore, it has been shown that
while point-group symmetry is not necessary, it can sometimes be
sufficient to guarantee the existence of isolated band touchings at
points of symmetry~\cite{manes-prb12}.

In this paper, we classify the band crossings occurring on the
symmetry lines of 3D crystals with screw rotational symmetry and
broken $\P*\T$ symmetry. We first describe the types of crossings that
are possible at generic points along a symmetry line. We then
specialize to $\T$-invariant, $\P$-broken crystals and consider the
crossings at $\T$-invariant points on those lines, for both spinful
and spinless $\T$ symmetry; the former applies to electrons in
crystals, and the latter to the spectrum of photonic
crystals~\cite{lu-natphot14}, as well as to electronic bands
calculated without including spin-orbit coupling.  

We find, for example, that in nonmagnetic crystals with a 3-fold screw
axis Weyl nodes are unavoidable at the symmetry points $\Gamma$ and A
when spin-orbit is included.  In a crystal like Te where the band
structure is composed of 6-band complexes, each complex generates a
triple Weyl node with $\chi=\pm 3$ at $\Gamma$ and another at A (in
addition to two single Weyl nodes with $\chi=\pm 1$ at each of those
points).  In contrast, the occurrence of triple nodes at generic
points along a symmetry axis requires 6-fold
symmetry~\cite{fang-prl12}.  The off-axis splitting of the bands in
the orthogonal plane is qualitatively different in the two cases: It
is cubic when the triple node is pinned to a $\T$-invariant point
(either $\Gamma$ or~A) on a 3-fold or 6-fold axis, and quadratic when
the triple node occurs at a generic point along a 6-fold axis. Thus, a
quadratic in-plane band splitting does not necessarily mean that the
Weyl node is a double node.

The article is organized as follows.  In Sec.~\ref{sec:rot-symm} we
classify the Weyl nodes occurring at generic points along a rotation
axis in the BZ. That section follows closely the discussion in
Ref.~\cite{fang-prl12} which we extend from pure rotations to screw
rotations, and it also includes a new result on the in-plane
dispersion of triple nodes. In. Sec.~\ref{sec:cobalt} we apply the
classification scheme to Weyl nodes on the 6-fold axis in the BZ of
ferromagnetic hexagonal close-packed (hcp) Co.  In
Sec.~\ref{sec:timerev-symm} we turn to nonmagnetic acentric crystals
and classify the degeneracies occurring at $\T$-invariant momenta on a
rotation axis.  As an example, we study in Sec.~\ref{sec:tellurium}
the Weyl nodes on the 3-fold axis in the BZ of trigonal Te. In
Sec.~\ref{sec:trbreak} we study the effect of a perturbation that
breaks $\T$ symmetry but preserves a 3-fold or 6-fold screw symmetry,
on the examples of Te and NbSi$_2$ respectively.  The conclusions are
drawn in Sec.~\ref{sec:conclusions}, and some supplementary
information and derivations are given in the appendices.

\section{Weyl nodes at generic points along a rotation axis}
\label{sec:rot-symm}

In this section we consider the most general scenario in which Weyl
points can occur along a symmetry line. Since their presence anywhere
in the BZ requires broken $\P*\T$ symmetry, we assume this to be the
case for our crystal.  Examples include ferromagnetic metals such as
body-centered cubic Fe~\cite{gosalbez-prb15} and hcp Co, nonmagnetic
acentric semiconductors such as trigonal
Te~\cite{hirayama-prl15,nakayama-prb17}, and polar conductors such as
TaAs, a Weyl semimetal~\cite{weng-prx15,huang-natcomm15}.  In the
first two examples $\T$ symmetry is broken and $\P$ symmetry is
present, while the reverse is true for the others.  Note that certain
antiferromagnets such as Cr$_2$O$_3$ do not qualify: They break $\P$
and $\T$ individually, but respect $\P*\T$.

We further assume that our crystal is left invariant under either a
pure rotation or a screw operation $n'_{m'}$, where $n'=2,3,4,6$
denotes a counterclockwise $2\pi/n'$ rotation around the $+\zhat$
axis, and the non-negative integer $m' < n'$ indicates a translation
along $+\zhat$ by a fraction $m'/n'$ of the lattice constant $c$
(which we take as the unit of length).  If $n$ is a divisor of $n'$,
invariance under $n'_{m'}$ implies invariance under $n_m$ about the
same axis, where $m=m'$ mod~$n$.  In the presence of $n_m$ symmetry
the Bloch Hamiltonian $H_{ij}(\kk)=\me{\psi_{i\kk}}{H}{\psi_{j\kk}}$
satisfies
\beq
\label{eq:rot-symm}
C_{n_m}H(\kk)C_{n_m}^{-1}=H(R_n\kk),
\eeq
where the matrix $C_{n_m}$ represents the $n_m$ operation in the Bloch
basis and $R_n\kk$ is the vector obtained by applying a
counterclockwise rotation of $2\pi/n$ to $\kk$. In writing
\eq{rot-symm} we have adopted the ``active picture'' where the action
of a transformation $S$ on a function $f(\rr)$ is described by
$Sf(\rr)=f(S^{-1}\rr)$.

We want to study the possible crossing between two eigenstates
$\ket{u}$ and $\ket{v}$ of $H(\kk)$ along a rotationally-invariant
line (a line where $R_n\KK=\KK$ mod~$\GG$ at every point $\KK$).  For
clarity we will focus on the axis
\beq
\label{eq:k-axis}
\KK=(0,0,\kz)
\eeq
that has the highest rotational symmetry $n=n'$, but our analysis also
applies to the other invariant lines that are present in the BZ as a
result of lattice periodicity.  Henceforth we will use the symbol
$\KK$ to refer to a point with coordinates given by \eq{k-axis}.

If the two states are very close in energy at $\KK$ and comparatively
far from other bands, we can work in the basis $\ket{u}=(1,0)^T$ and
$\ket{v}=(0,1)^T$ choosing $\ket{u}$ as the higher-energy state at
$K_z+\delta$ when $\delta\rightarrow 0^+$, and approximate the Bloch
Hamiltonian around $\KK$ by
\beq
\label{eq:ham-eff}
H_{\rm eff}(\KK+\qq)=d(\qq)\mathbbm{1}+
f(\qq)\sigma_+ +f^*(\qq)\sigma_-+g(\qq)\sigma_z, 
\eeq
where $\qq=(q_x,q_y,q_z)$, $\mathbbm{1}$ is the $2\times 2$ identity
matrix, $\sigma_\pm=\sigma_x\pm i\sigma_y$, and a dependence on $\KK$
of the real functions $d$ and $g$ and of the complex function $f$ is
implied. In this approximation the two basis states are eigenstates of
$H_{\rm eff}(\KK)$, which means that $f(\qq=0)=0$.  The condition for
a crossing to occur at $\KK$ is that $g(\qq=0)=0$ as well, and in the
following it is assumed we have found such a point. The functions $f$
and $g$ can then be expanded around $\KK$ as
\begin{subequations}
\label{eq:expansion}
\begin{align}
\label{eq:f-expansion}
f(q_+,q_-,q_z)&=\sum_{n_1n_2n_3}\,A_{n_1n_2n_3}
q_+^{n_1}q_-^{n_2}q_z^{n_3},\\
\label{eq:g-expansion}
g(q_+,q_-,q_z)&=\sum_{m_1m_2m_3}\,B_{m_1m_2m_3}
q_+^{m_1}q_-^{m_2}q_z^{m_3},
\end{align}
\end{subequations}
where $q_\pm=q_x\pm iq_y$, $A_{n_1n_2n_3}$ and $B_{m_1m_2m_3}$ are
$\kz$-dependent complex coefficients with $n_i,m_i\geq 0$, and
$A_{000}=B_{000}=0$ by assumption.
The requirement that \eq{ham-eff} be Hermitian implies that $g(\qq)$
is real, leading to the relation
\beq 
\label{eq:B-cond}
B_{m_1m_2m_3}=B^*_{m_2m_1m_3}.  
\eeq

The types of crossings that can occur at generic points along the
axis, where $n_m$ symmetry is present but $\T$ is broken, were
classified in Ref.~\cite{fang-prl12} by imposing the rotational
constraint~(\ref{eq:rot-symm}) on the effective Hamiltonian of
\eqs{ham-eff}{expansion}. The authors specialized to pure $n$-fold
rotations ($m=0$), and in the following we extend their treatment to
include screw rotations. As we shall see, the resulting classification
is independent of the pitch of the screw.

At any point along the axis \eq{rot-symm} reduces to
$[C_{n_m},H(\KK)]=0$, so that the energy eigenstates are also
eigenstates of $C_{n_m}$.  The rotational eigenvalues~$\alpha(\kz)$
are determined by noting that $(n_m)^n$ describes a $2\pi$ rotation
around the $+\zhat$ axis followed (or preceeded) by a translation by
$m\zhat$.  The former leaves a spinless wave function unchanged but
flips the sign of a spinful wave function, and the latter gives an
extra phase factor $e^{-im\kz}$ (the minus sign comes from using the
active picture). Taking the $n$-th roots we find
\begin{subequations}
\label{eq:rot-eig}
\begin{align}
\alpha_p(\kz)&=\gamma_{n_m}(\kz)\,e^{i2\pi p/n},\\
\gamma_{n_m}(\kz)&=e^{i\pi (F-F'm)/n},
\end{align}
\end{subequations}
where $p$ is an integer chosen between~0 and $n-1$, $F=0$ ($F=1$) for
spinless (spinful) $\T$ symmetry, and $F'=\kz/\pi$. In the same basis
of \eq{ham-eff}, then, the matrix describing the $n_m$ operation reads
\beq
\label{eq:screw-eff}
C_{n_m}(\kz)=\gamma_{n_m}(\kz) 
\left(
  \begin{array}{cc}
    e^{i2\pi p_u/n} & 0\\
    0 & e^{i2\pi p_v/n}
  \end{array}
\right).
\eeq
Combining Eqs.~(\ref{eq:rot-symm}), (\ref{eq:ham-eff}), and
(\ref{eq:screw-eff}) gives
\begin{subequations}
\begin{align}
\label{eq:f-rot}
e^{i\frac{2\pi}{n}(p_u-p_v)}f(q_+,q_-,q_z)
&=f\left(q_+e^{i\frac{2\pi}{n}},q_-e^{-i\frac{2\pi}{n}},q_z\right),\\
\label{eq:g-rot}
g(q_+,q_-,q_z)&
=g\left(q_+e^{i\frac{2\pi}{n}},q_-e^{-i\frac{2\pi}{n}},q_z\right).
\end{align}
\end{subequations}
Inserting \eq{f-expansion} in \eq{f-rot} and \eq{g-expansion} in
\eq{g-rot} we find
\begin{subequations}
\begin{eqnarray}
\label{eq:nnpp}
n_1-n_2 =&\;p_u-p_v \quad &\hbox{mod} \;\; n,\\
\label{eq:mm}
m_1-m_2 =&0 \quad &\hbox{mod} \;\; n.
\end{eqnarray}
\end{subequations}
The only nonzero elements of $A_{n_1n_2n_3}$ or $B_{m_1m_2m_3}$ occur
when \eq{nnpp} or~(\ref{eq:mm}) is satisfied, respectively. When
$p_u=p_v$ the degeneracy is nonchiral and can be gapped by a small
perturbation that respects $n_m$ symmetry (see
Appendix~\ref{app:level-repulsion}), whereas for $p_u\not=p_v$ it is
chiral and robust.

We will use two criteria to classify the Weyl nodes that occur for
$p_u\ne p_v$: (i) the power laws that describe at leading order the
splitting of the bands as one moves away from the node along the axis
and in the orthogonal directions, and (ii) the chiral charge of the
node. 

Let us consider first the splitting of the bands. Regarding the
behavior along the axis, the only surviving terms in
\eqs{f-expansion}{g-expansion} when $q_+=q_-=0$ are those with
$n_1=n_2=0$ and $m_1=m_2=0$ respectively. Then $A_{00n_3}$ vanishes
for all $n_3$ because of \eq{nnpp}, and to leading order in $q_z$ we
find
\beq
\label{eq:ham-eff-parallel}
H_{\rm eff}(0,0,\kz+q_z)=B_{001}\sigma_z q_z.
\eeq
From this we conclude that the band splitting is generically linear
along the symmetry axis. With our choice of $\ket{u}=(1,0)^T$ as the
higher-energy state on the higher-$\kz$ side of the crossing,
$B_{001}$ is positive.\footnote{In Ref.~\cite{fang-prl12}, the
  opposite choice was made, i.e., $B_{001}<0$ in
  \eq{ham-eff-parallel}. As a result, the chiral charges in Table~I of
  that work have the opposite signs compared to our
  Table~\ref{table:generic}.}

The behavior on the orthogonal plane is described by
$H_{\rm eff}(q_x,q_y,K_z)$.  We now need to collect the leading terms
with $n_3=0$ in \eq{f-expansion} that satisfy
condition~(\ref{eq:nnpp}). Those terms determine the magnitude of the
chiral charge, and their form only depends on $n$ and on the ratio
\beq
\label{eq:rot-ratio}
\frac{\alpha_u}{\alpha_v}=e^{i2\pi(p_u-p_v)/n}
\eeq
between the rotational eigenvalues of the crossing states.  We must
also collect terms with $m_3=0$ in \eq{g-expansion} that comply with
condition~(\ref{eq:mm}). At leading order we find
$g(q_+,q_-,0)=B_{110}q_+q_-$, or equivalently,
\beq
\label{eq:g-nonchiral}
g(q_x,q_y,0)=B_{110}(q_x^2+q_y^2).  
\eeq
This term is allowed for all $n$, and it appears to have been
overlooked in Ref.~\cite{fang-prl12}.  It does not affect the sign or
magnitude of the chiral charge, but in some cases it qualitatively
changes the off-axis band splitting.  For triple nodes, in particular,
this term dominates the band splitting in the plane normal to the
axis, although a cubic splitting would still be evident if one could
follow the parabolic $g(\qq)\!=\!0$ surface instead of the fixed-$q_z$
plane.

\begin{table}
  \caption{\label{table:generic} Classification of Weyl nodes
    at generic points on an $n$-fold axis 
    in the BZ
    of a crystal
    with $n_m$ symmetry and broken $\P*\T$ symmetry. 
    $\alpha_u$ and $\alpha_v$ are the rotational eigenvalues
    of the crossing states, with $u$ denoting the higher-energy state
    on the higher-$\kz$ side of the crossing.
    The Hamiltonian near a node on the  plane
    perpendicular to the axis is $H_{\rm eff}(q_x,q_y,K_z)=
    h_{\rm eff}+h^\dagger_{\rm eff}$, and
    $q_\pm=q_x\pm iq_y$. 
    In each row the complex parameters $a$ and $b$ correspond to 
    specific coefficients $A_{n_1n_20}$ in \eq{f-expansion},
    and $c=B_{110}/2$
    from \eq{g-nonchiral} is real (the term $c(q_x^2+q_y^2)$ is  written only when it is of leading order).
    $\chi$~is the chiral charge with $\chi_{ab}=\text{sgn}(|b|-|a|)$,
    and
    the off-axis band splitting at leading order is indicated as $q_\perp$ (linear) or
    $q_\perp^2$ (quadratic).
  }
\begin{tabular}{cccccc}
  $n$ & $\alpha_u/\alpha_v$ & $h_{\rm eff}$ & $\chi$ & Splitting\\
  \hline\hline
  2 & -1& $\left(aq_++bq_-\right)\sigma_+$ &$\chi_{ab}$ & $q_\perp$\\
  \hline
  3 & $e^{\pm i2\pi/3}$ & $aq_\pm\sigma_+$ & $\mp 1$ & $q_\perp$ \\
  \hline
  4 & $\pm i$ & $aq_\pm\sigma_+$ & $\mp 1$ & $q_\perp$ \\
  4 & -1& $\left(aq_+^2+bq_-^2\right)\sigma_++c\left( q_x^2+q_y^2\right)\sigma_z$ &$2\chi_{ab}$ & $q_\perp^2$\\
  \hline
  6 & $e^{\pm i\pi/3}$ &$aq_\pm\sigma_+$ & $\mp 1$ & $q_\perp$ \\
  6 & $e^{\pm 2i\pi/3}$ &$aq_\pm^2\sigma_++c\left( q_x^2+q_y^2\right)\sigma_z$ & $\mp 2$ & $q_\perp^2$ \\
  6 & $-1$ &$\left(aq_+^3+bq_-^3\right)\sigma_++c\left( q_x^2+q_y^2\right)\sigma_z$ & $3\chi_{ab}$ & $q_\perp^2$ \\
\end{tabular}
\end{table}

By now we have gathered all the needed information to catalog the Weyl
crossings that can occur at generic points on a rotation axis. The
classification is given in Table~\ref{table:generic}, and as
anticipated it does not depend on the pitch $m/n$ of the screw.  The
main conclusions are as follows. The occurrence of triple Weyl nodes
requires 6-fold symmetry, double nodes require 4-fold or 6-fold
symmetry, and rotation axes of any order can host single Weyl nodes.
At leading order the band splitting along the axis is linear in all
cases, while on the orthogonal plane it is linear for single nodes and
quadratic for both double and triple nodes. These conclusions agree
with Ref.~\cite{fang-prl12}, except for the realization that the
in-plane splitting of a triple Weyl node is generally quadratic, not
cubic. In Sec.~\ref{sec:timerev-symm}, we will encounter triple nodes
for which the quadratic term~(\ref{eq:g-nonchiral}) is disallowed by
symmetry, resulting in a cubic in-plane dispersion.

\section{Application to hcp Cobalt}
\label{sec:cobalt}

In order to illustrate the preceding discussion, we have performed an
{\it ab initio} study of the band structure of hcp Co. The technical
details of the calculation are given in Appendix~\ref{app:methods}.
In the hcp structure (space group P6$_3$/mmc, No. 194), the $c$ axis
is a $6_3$ screw axis. This is a ``neutral screw'' (a screw that has
neither right or left sense~\cite{buerger-book56}) that coexists with
$\P$ symmetry, while $\T$ is broken by the ferromagnetic order. The
spontaneous breaking of $\T$ symmetry occurs in the spin channel via
the exchange interaction, and is then transmitted to the orbital wave
functions and to the band structure by the spin-orbit interaction. In
our calculation, the magnetization points along the positive haxagonal
axis.
 
\begin{figure}
\begin{center}
\includegraphics[width=1.0\columnwidth]{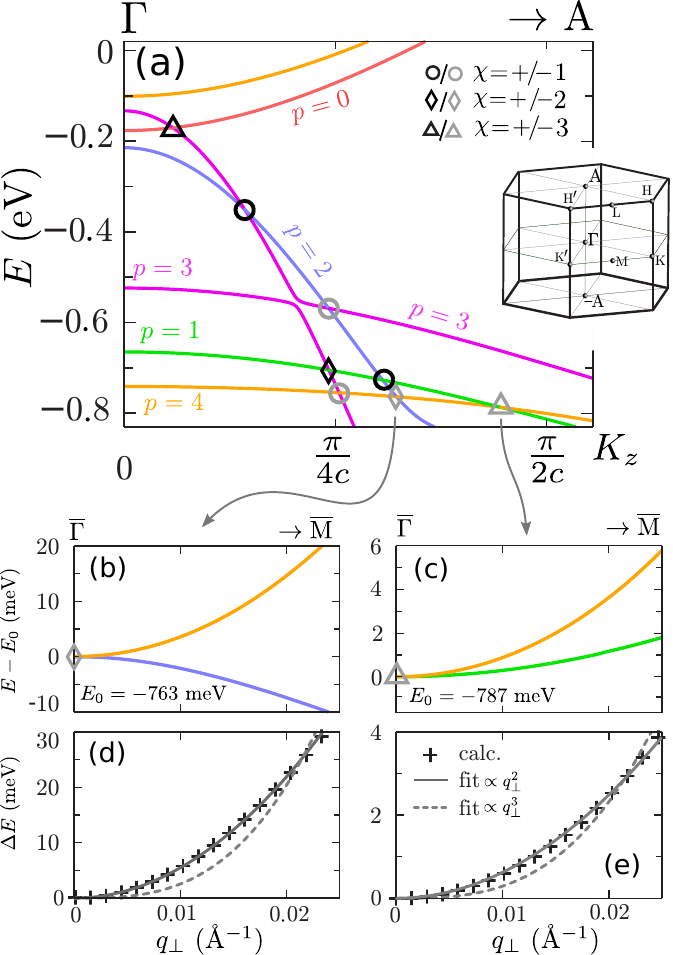}
\end{center}
\caption{\label{fig1} (Color online) (a) Calculated band structure of
  hcp Co along the sixfold symmetry line $\Gamma$A.  Energies are
  measured from the Fermi level, and each color denotes a branch
  labeled by the integer $p$ in \eq{rot-eig}.  Markers denote Weyl
  crossings with the indicated chiral charges $\chi$.  The inset shows
  the hexagonal BZ and its high-symmetry points.  (b) and (c) show the
  in-plane dispersions near a double and a triple Weyl node
  respectively, and (d) and (e) show the corresponding band splittings
  away from the nodes, together with the quadratic and cubic best
  fits.}
\end{figure}

Figure~\ref{fig1}(a) shows the energy bands near the Fermi level on a
segment of the 6-fold axis $\Gamma$A. (For a more complete picture of
the band structure of Co, see Ref.~\cite{McMullan92}.)  The different
branches are color-coded by the rotational labels $p$ in~\eq{rot-eig},
which were determined directly from the Bloch wave functions. Already
in this narrow energy range of $\sim 0.8$~eV one can find all the
types of crossings listed in Table~\ref{table:generic} for $n=6$.  For
example, the crossing between the $p_u=4$ and $p_v=2$ branches near
the bottom of the figure is a double Weyl node of negative chirality
(chiral charge $\chi=-2$) because $\alpha_u/\alpha_v=e^{i2\pi/3}$,
while the two crossings with $(p_u,p_v)=(2,3)$ and $(3,2)$ have
$\chi=+1$ and $\chi=-1$, respectively. As expected, the crossing
between the two branches with $p=3$ is avoided.  Triple Weyl nodes are
generated at the crossings where $p_u-p_v=3\;\mathrm{mod}\;6$, namely,
$(p_u,p_v)=(0,3)$ and~$(4,1)$; in this case the chirality
${\rm sgn}(\chi)$ cannot be extracted from the symmetry labels.  For
each node, we have evaluated $\chi$ explicitly from the quantized
Berry-curvature flux through a small enclosing box, as described in
Appendix~\ref{app:methods}.  For the single and double nodes, the
calculated values of $\chi$ agree in sign and magnitude with those
predicted from the symmetry labels.
  
Figure~\ref{fig1}(a) confirms that the band dispersions along the axis
are linear around every Weyl node.  The dispersions are also linear in
the transverse directions when $\vert\chi\vert=1$ (not shown), but not
when $\vert\chi\vert>1$.  Figures~\ref{fig1}(b) and~(c) show the
dispersions near a double and a triple node respectively, along the
in-plane direction $\overline{\Gamma\mathrm{M}}$ (denoted as
$q_\perp$) on the constant-$k_z$ plane of the node; in both cases, the
in-plane dispersion is fairly isotropic near
$q_\perp=0$. Figures~\ref{fig1}(d) and (e) show the band splittings
with increasing $q_\perp$, together with their best fits by quadratic
and cubic functions.  The splitting at small $q_\perp$ is accurately
described by $\propto q_\perp^2$ (rather than $\propto q_\perp^3$) for
both the double {\it and} the triple node.  This confirms that the
in-plane dispersion of a triple node is dominated by the nonchiral
quadratic term in \eq{g-nonchiral}, which masks the cubic dispersion
from the chiral term $\left( A_{300}q_+^3+A_{030}q_-^3\right)\sigma_+$
in the last row of Table~\ref{table:generic}.

We are aware of one other work~\cite{chen-natcomm16} where triple Weyl
nodes at generic points on a 6-fold axis were studied numerically (for
a haxagonal photonic crystal). That work only reports the band
dispersions along the axis, which as expected are linear around each
node (see Fig.~4 therein). The authors state that the in-plane
dispersion is cubic for the triple nodes, but it is unclear whether
this was verified numerically, or if it is simply a remark based on
the conclusions of Ref.~\cite{fang-prl12}.

\section{Weyl nodes at time-reversal invariant momenta on a rotation
  axis}
\label{sec:timerev-symm}

Let us resume the formal development of our systematic classification.
In Sec.~\ref{sec:rot-symm} we considered a broken-$\P*\T$ crystal with
$n_m$ symmetry, and classified the Weyl nodes occurring at generic
points on the rotation axis of \eq{k-axis}.  For the remainder of this
work we specialize to a $\T$-invariant, $\P$-broken crystal, and focus
on points along that axis where $\T$ symmetry is present (the
time-reversal invariant momenta, or TRIM).  Since $\T$ maps $\kk$ onto
$-\kk$ mod $\GG$, such points occur at $\kz=0$ and $\pi$,
corresponding to $F'=0$ and $1$ respectively in \eq{rot-eig}. As for
the additional symmetry lines not passing through~$\Gamma$, they only
contain TRIM if they are 2-fold or 4-fold axes~\footnote{Consider a
  crystal with $n_m$ symmetry. For $n=3$ the TRIM that do not project
  onto $\overline{\Gamma}$ on the projected 2D BZ are at the three
  $\overline{\rm M}$ points, while the 3-fold axes in $k$ space are at
  $\overline{\rm K}$ and $\overline{\rm K'}$, so there are no TRIM on
  a 3-fold axis except at $\overline{\Gamma}$.  For $n=6$ there are no
  6-fold axes off $\overline{\Gamma}$, only 3-fold axes at
  $\overline{\rm K}$ and $\overline{\rm K'}$ which do not contain
  TRIM, and 2-fold axes at $\overline{\rm M}$, $\overline{\rm M}'$,
  and $\overline{\rm M}''$.  Thus, 3-fold and 6-fold axes never
  contain TRIM except at $\overline{\Gamma}$.}. The ensuing analysis
assumes that $n$-fold rotational symmetry and $\T$ are the only
symmetries present at the TRIM.

\subsection{Formal derivation}
\label{sec:timerev-symm-formal}

We now find it more convenient to identify an eigenvalue of $C_{n_m}$
by an index $j$ via [compare with \eq{rot-eig}]
\beq
\label{eq:rot-eig-trim}
\alpha_j=e^{i\pi j/n}\quad\text{where}\quad j=2p+F-F'm,
\eeq
so that the $j$ are integers spaced two units apart.  With this
notation, $\T$ maps $\alpha_j$ into $\alpha_{-j}$.  If the two
time-reversed states $\ket{u}$ and $\ket{v}=\T\ket{u}$ are distinct,
they form a degenerate pair~\footnote{The degeneracy follows from
  \eq{T-symm}, which at the TRIM becomes $[\T,H(\kk)]=0$.}  to which
we assign the labels $j_u=j$ and $j_v=-j$.  We then wish to understand
whether and how this pair splits as we move off the symmetry axis as
described by Eqs.~(\ref{eq:ham-eff}) and (\ref{eq:f-expansion}), and
to determine the monopole charge when the crossing is chiral.  In this
context, the results of the rotational symmetry analysis in
Sec.~\ref{sec:rot-symm} can be expressed via $p_u-p_v=(j_u-j_v)/2=j$,
so that the only nonzero elements of $A_{n_1n_2n_3}$ occur when
\beq
n_1-n_2= j \quad \hbox{mod} \; n ,
\label{eq:nnj}
\eeq
while nonzero elements of $B_{m_1m_2m_3}$ must still comply with
condition~(\ref{eq:mm}).

Let us now turn to the conditions imposed by $\T$ itself, which are
different depending on whether $F=0$ or 1 because $\T^2=(-1)^F$.  The
first thing to note is that unless the two time-reversed states have
different rotational labels, when $F=0$ they are actually the same
state and there is no $\T$-protected degeneracy.  Thus, we impose
\beq
\label{eq:no-degen}
j\not= 0,n\quad\text{when}\quad F=0.
\eeq
(From here on, $j$ is chosen between 0 and $2n-1$.)  When $F=1$ the
Kramers theorem guarantees that the two states are
different~\cite{sakurai-book94} and the above restriction does not
apply.  These conclusions are in line with the ``Wigner rules'' for
degeneracies with $\T$ symmetry present~\cite{heine-book60}.

$\T$~symmetry also imposes the restrictions
\begin{subequations}
\begin{eqnarray}
\label{eq:T-cond-f-a}
n_1+n_2+n_3&=\;F&\text{ mod }2,\\
\label{eq:T-cond-g}
m_1+m_2+m_3&=\;	1&\text{ mod }2
\end{eqnarray}
\end{subequations}
on the nonzero elements of $A_{n_1n_2n_3}$ and $B_{m_1m_2m_3}$,
respectively, as a result of the relation
\beq
\label{eq:T-symm}
\T H(\kk)\T^{-1}=H(-\kk).
\eeq
To show this, let us express $\T$ in our basis by acting with it on a
state $\ket{w}=a\ket{u}+b\ket{v}$.  From $\T\ket{v} =(-1)^F\ket{u}$
and the antilinearity of $\T$ we obtain $\T=\sigma_x{\cal K}$ ($F=0$)
and $\T=-i\sigma_y{\cal K}$ ($F=1$) with ${\cal K}$ the complex
conjugation operator, so that $\T f(\qq)\T^{-1}=f^*(\qq)$ and
$\T\sigma_\pm\T^{-1}=(-1)^F\sigma_\mp$.  Inserting \eq{ham-eff} in
\eq{T-symm} and using these identities gives $f(-\qq)=(-1)^Ff(\qq)$,
which leads to \eq{T-cond-f-a} when combined with \eq{f-expansion}.
Similarly, from $\T g(\qq)\T^{-1}=g(\qq)$ and
$\T\sigma_z\T^{-1}=-\sigma_z$ we get $g(-\qq)=-g(\qq)$, which leads to
\eq{T-cond-g}.

Armed with the above relations, we can proceed to classify the
degeneracies at the TRIM $(0,0,0)$ and $(0,0,\pi)$.  We begin with the
on-axis band splittings. As in Sec.~\ref{sec:rot-symm}, we collect
terms in \eq{g-expansion} with $m_1=m_2=0$, excluding $B_{000}$ which
vanishes by assumption. \equ{mm} is automatically satisfied and
\eq{T-cond-g} forces $m_3$ to be odd, and so the leading term is
generically
\beq
\label{eq:g-00qz}
g(0,0,q_z)=B_{001}q_z.
\eeq
Turning to the expansion~(\ref{eq:f-expansion}) of $f(0,0,q_z)$, we
keep terms with $n_1=n_2=0$ excluding $A_{000}$. \equ{nnj} requires
$j=0$ or~$n$ which conflicts with \eq{no-degen} when $F=0$, and when
$F=1$ \eq{T-cond-f-a} requires $n_3$ to be odd.  Thus
\beq
\label{eq:f-z}
f(0,0,q_z)=
\begin{cases}
A_{001}q_z,
&\text{when $F=1$ and $j=0$ or~$n$}\\
0,&\text{otherwise}.
\end{cases}
\eeq
In all cases $H_{\rm eff}(0,0,\kz+q_z)$ is linear in $q_z$, producing
a linear band splitting along the axis.

In order to describe the in-plane behavior, let us collect the leading
terms with $n_3=0$ in \eq{f-expansion}. \equ{T-cond-f-a} can then be
written as
\beq
\label{eq:T-cond}
n_1-n_2=F\text{ mod }2,
\eeq
which together with \eqs{nnj}{no-degen} constrains the form of
$f(q_+,q_-,0)$.  Turning to $g(q_+,q_-,0)$ and setting $m_3=0$ in
\eq{T-cond-g} we conclude that $m_1-m_2$ must be odd, which excludes
terms with $m_1=m_2$ such as \eq{g-nonchiral}.  The requirement that
$m_1-m_2$ be odd conflicts with condition~(\ref{eq:mm}) when $n$ is
even, and so we find
\beq
\label{eq:g-off-axis-trim}
g(q_+,q_-,0)=
\begin{cases}
2{\rm Re}
\left(B_{300}q_+^3\right),&\text{for $n=3$}\\
0,&\text{for $n=2,4,6$}.
\end{cases}
\eeq

In summary, the band splitting moving away from a degeneracy protected
by $\T$ and $n_m$ symmetry at $\kz=0$ or $\pi$ is generically linear
along the axis.  Assuming no other symmetries, the form of the
in-plane Hamiltonian $H_{\rm eff}(q_+,q_-,\kz)$ is constrained by
Eqs.~(\ref{eq:nnj}), (\ref{eq:no-degen}), (\ref{eq:T-cond}),
and~(\ref{eq:g-off-axis-trim}); the type $F$ of $\T$ symmetry enters
the first three equations, and an additional dependence on the pitch
$m/n$ is introduced by \eqs{nnj}{no-degen} at $\kz=\pi$.  The
Weyl-like solutions compatible with lattice periodicity are listed in
Table~\ref{table:spinless} for spinless $\T$, and in
Table~\ref{table:spinful} for spinful $\T$.  With spinless $\T$ all
Weyl degeneracies are double nodes, and with spinful $\T$ they are
either single or triple nodes.  Triple nodes occur not only for $n=6$
as in Table~\ref{table:generic} but also for $n=3$, and in both cases
the in-plane splitting is cubic, not quadratic as in
Table~\ref{table:generic}.

\begin{table}
  \caption{\label{table:spinless} Classification of spinless Weyl nodes
    at the TRIM  $\kz=0$ and $\pi$ on an $n$-fold axis
    in the BZ of a crystal with $n_m$ and $\T$ symmetries.
    Each row is uniquely identified by the values of $n$ and of the symmetry label $j$
    [defined in \eq{rot-eig-trim}] of the higher-energy state on the higher-$\kz$ side of the crossing; from 
    these two values,
    the entries in the
    remaining columns can be generated. The values of $m$
    for which the given $j$ occurs at $\kz=0$ and $\pi$
    are listed under $m(0)$ and $m(\pi)$, respectively, and
    the rest of the notation follows Table~\ref{table:generic}.}
\begin{tabular}{ccccccc}
$n$ & $m(0)$  & $m(\pi)$ & $j$ & $h_{\rm eff}$ & $\chi$ & Splitting\\
\hline\hline
3 & --- &  1   & 1 & $aq_-^2\sigma_+$ &$+2$ & $q_\perp^2$\\
3 & all &  0,2 & 2 & $aq_+^2\sigma_+$ &$-2$ & $q_\perp^2$\\
3 & all &  0,2 & 4 & $aq_-^2\sigma_+$ &$+2$ & $q_\perp^2$\\
3 & --- &  1   & 5 & $aq_+^2\sigma_+$ &$-2$ & $q_\perp^2$\\
\hline
4 & all & 0,2 & 2  & $\left(aq_+^2+bq_-^2\right)\sigma_+$ &$2\chi_{ab}$ & $q_\perp^2$\\
4 & all & 0,2 & 6  & $\left(aq_+^2+bq_-^2\right)\sigma_+$ &$2\chi_{ab}$ & $q_\perp^2$\\
\hline
6 & all&  0,2,4 & 2 & $aq_+^2\sigma_+$ &$-2$ & $q_\perp^2$\\
6 & all&  0,2,4 & 4 & $aq_-^2\sigma_+$ &$+2$ & $q_\perp^2$\\
6 & all&  0,2,4 & 8 & $aq_+^2\sigma_+$ &$-2$ & $q_\perp^2$\\
6 & all&  0,2,4 &10 & $aq_-^2\sigma_+$ &$+2$ & $q_\perp^2$
\end{tabular}
\end{table}

\begin{table}
  \caption{\label{table:spinful} 
    Classification of spinful Weyl nodes
    at the TRIM  $\kz=0$ and $\pi$ on an $n$-fold axis
    in the BZ of a crystal with $n_m$ and $\T$ symmetries.
    The notation
    is the same as in Tables~\ref{table:generic} and \ref{table:spinless},
    except that here $c=B_{300}$ is a complex coefficient.}
\begin{tabular}{ccccccc}
$n$ & $m(0)$ & $m(\pi)$ &$j$ & $h_{\rm eff}$ & $\chi$ & Splitting\\
\hline\hline
2 & all & 0 & 1 & $aq_-\sigma_+$ &$+1$ & $q_\perp$\\
2 & all & 0 & 3 & $aq_+\sigma_+$ &$-1$ & $q_\perp$\\
\hline
3 & --- & 1 & 0 & $\left(aq_+^3+bq_-^3\right)\sigma_++cq_+^3\sigma_z$ &$3\chi_{ab}$ & $q_\perp^3$\\
3 & all & 0,2 & 1 & $aq_+\sigma_+$ &$-1$ & $q_\perp$\\
3 & --- & 1 & 2 & $aq_-\sigma_+$ &$+1$ & $q_\perp$\\
3 & all & 0,2  & 3 & $\left(aq_+^3+bq_-^3\right)\sigma_++cq_+^3\sigma_z$ &$3\chi_{ab}$ & $q_\perp^3$\\
3 & --- & 1 & 4 & $aq_+\sigma_+$ &$-1$ & $q_\perp$\\
3 & all & 0,2  & 5 & $aq_-\sigma_+$ &$+1$ & $q_\perp$\\
\hline
4 & all & 0,2 & 1 & $aq_+\sigma_+$ &$-1$ & $q_\perp$\\
4 & all & 0,2 & 3 & $aq_-\sigma_+$ &$+1$ & $q_\perp$\\ 
4 & all & 0,2 & 5 & $aq_+\sigma_+$ &$-1$ & $q_\perp$\\ 
4 & all & 0,2 & 7 & $aq_-\sigma_+$ &$+1$ & $q_\perp$\\
\hline
6 & all & 0,2,4& 1 & $aq_+\sigma_+$ &$-1$ & $q_\perp$\\
6 & all & 0,2,4& 3 & $\left(aq_+^3+bq_-^3\right)\sigma_+$ &$3\chi_{ab}$ & $q_\perp^3$\\
6 & all & 0,2,4& 5 &  $aq_-\sigma_+$ &$+1$ & $q_\perp$\\
6 & all & 0,2,4& 7 &  $aq_+\sigma_+$ &$-1$ & $q_\perp$\\
6 & all & 0,2,4& 9 & $\left(aq_+^3+bq_-^3\right)\sigma_+$ &$3\chi_{ab}$ & $q_\perp^3$\\
6 & all & 0,2,4&11 & $aq_-\sigma_+$ &$+1$ & $q_\perp$
\end{tabular}
\end{table}

\begin{figure}
\begin{center}
{\bf (a)} 
\includegraphics[width=2.75cm]{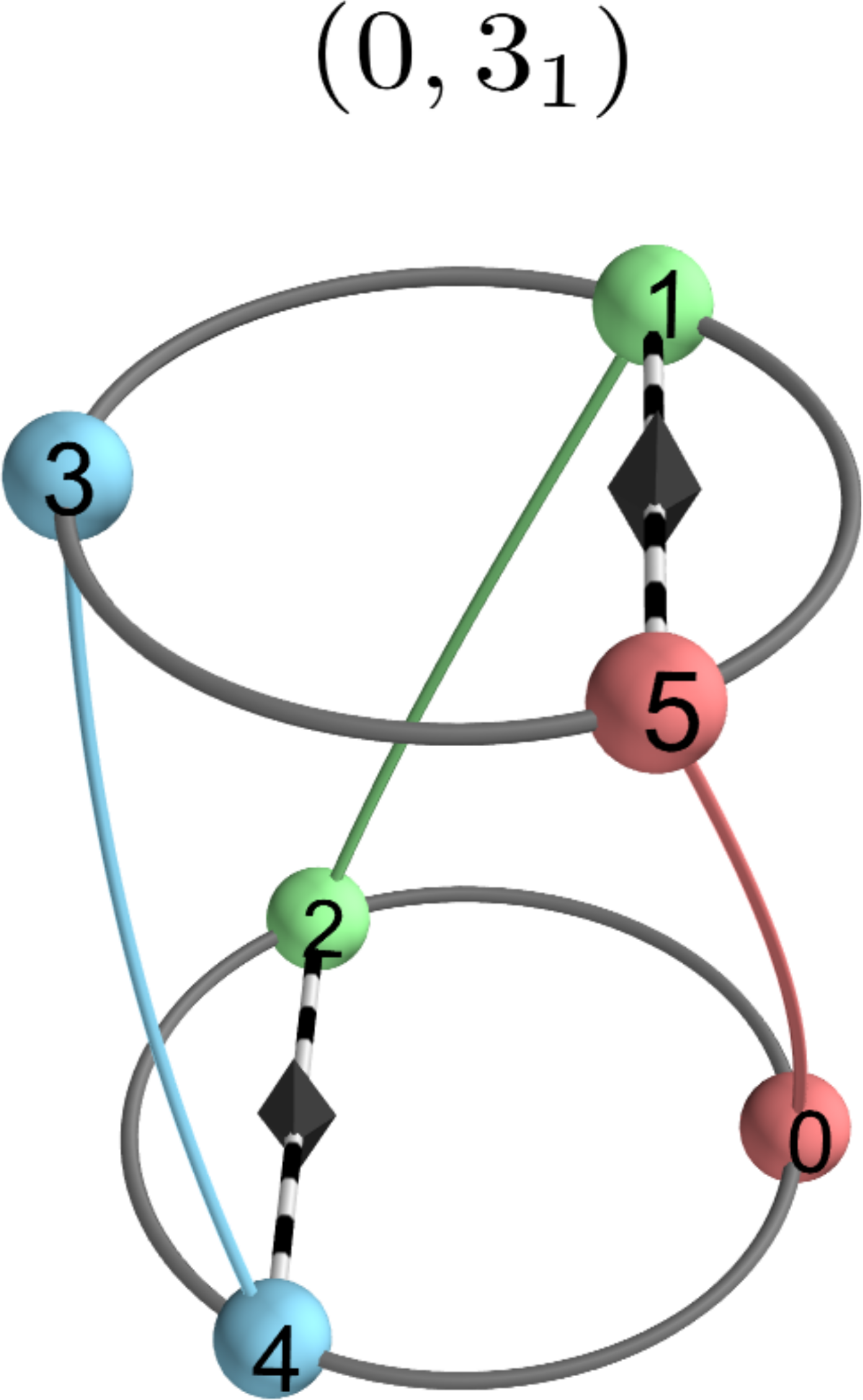}\quad
{\bf (b)} 
\includegraphics[width=4.1cm]{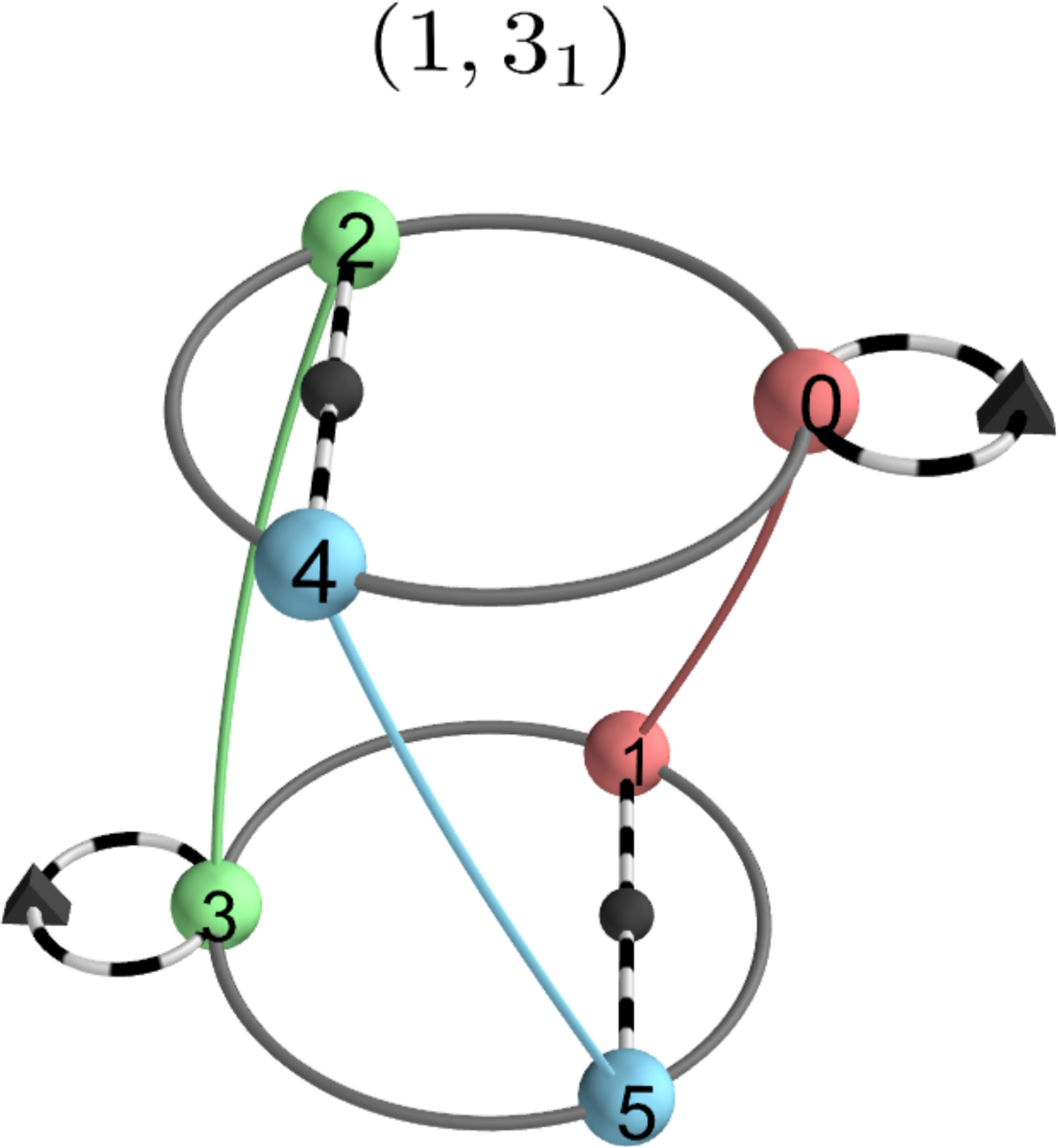}
\end{center}
\caption{\label{fig2} (Color online) Schematic representation of the
  degeneracies at the TRIM along a $3_1$-invariant axis.  (a) is
  for spinless $\T$ ($F=0$), and (b) for spinful $\T$ ($F=1$).
  Solid gray circles represent the complex unit circle, with the
  rotational eigenvalues of \eq{rot-eig-trim} at $\kz=0$ (bottom) and
  $\kz=\pi$ (top) marked as spheres labeled by $j$, and their windings
  with $\kz$ represented by lines with matching colors.  A dashed line
  or loop connecting spheres indicates that a pair of time-reversed
  states with those labels forms a single ($\CIRCLE$), double
  ($\blacklozenge$), or triple ($\blacktriangle$) Weyl node.}
\end{figure}

\subsection{Schematic description: $(F,n_m)$ diagrams}
\label{sec:timerev-symm-diagr}

Let us illustrate the use of Tables~\ref{table:spinless}
and~\ref{table:spinful} by considering some specific combinations
$(F,n_m)$. We start with $(0,3_1)$, i.e., spinless $\T$ and a
right-handed 3-fold screw. At $\kz=0$ along a $3_1$-invariant axis the
states carry labels $j=0,2,4$ [\eq{rot-eig-trim}].  The $j=0$ states
are nondegenerate [\eq{no-degen}], while time-reversed pairs of states
with labels $(j,-j+2n)=(2,4)$ or (4,2) form double Weyl nodes of
negative or positive chirality respectively (second and third rows of
Table~\ref{table:spinless}).  At $\kz=\pi$ the possible labels are
$j=1,3,5$; the $j=3$ states are nondegenerate, and the pairs (1,5) and
(5,1) form double Weyl nodes of positive and negative chirality
respectively.  As $\kz$ goes from $0$ to $\pi$ the rotational
eigenvalues of \eq{rot-eig-trim} wind as $e^{-i\kz/3}$, so that $j=0$
goes into $j=5$, $j=2$ into $j=1$, and $j=4$ into $j=3$.  Except for
the chiralities, all this information is presented schematically in
Fig.~\ref{fig2}(a).

Figure~\ref{fig2}(b) shows the $(1,3_1)$ diagram, where the allowed
$j$ values have shifted by $+1$ compared to Fig.~\ref{fig2}(a). With
spinful~$\T$ the restriction~(\ref{eq:no-degen}) does not apply, and
all bands pair up to form either single or triple Weyl nodes at both
$\kz=0$ and $\pi$.  The triple nodes are formed between Kramers pairs
with the same label $j=0$ or $j=3$ located on the equator of the unit
circle.

\begin{figure}
\begin{center}
{\bf (a)} 
\includegraphics[width=2.75cm]{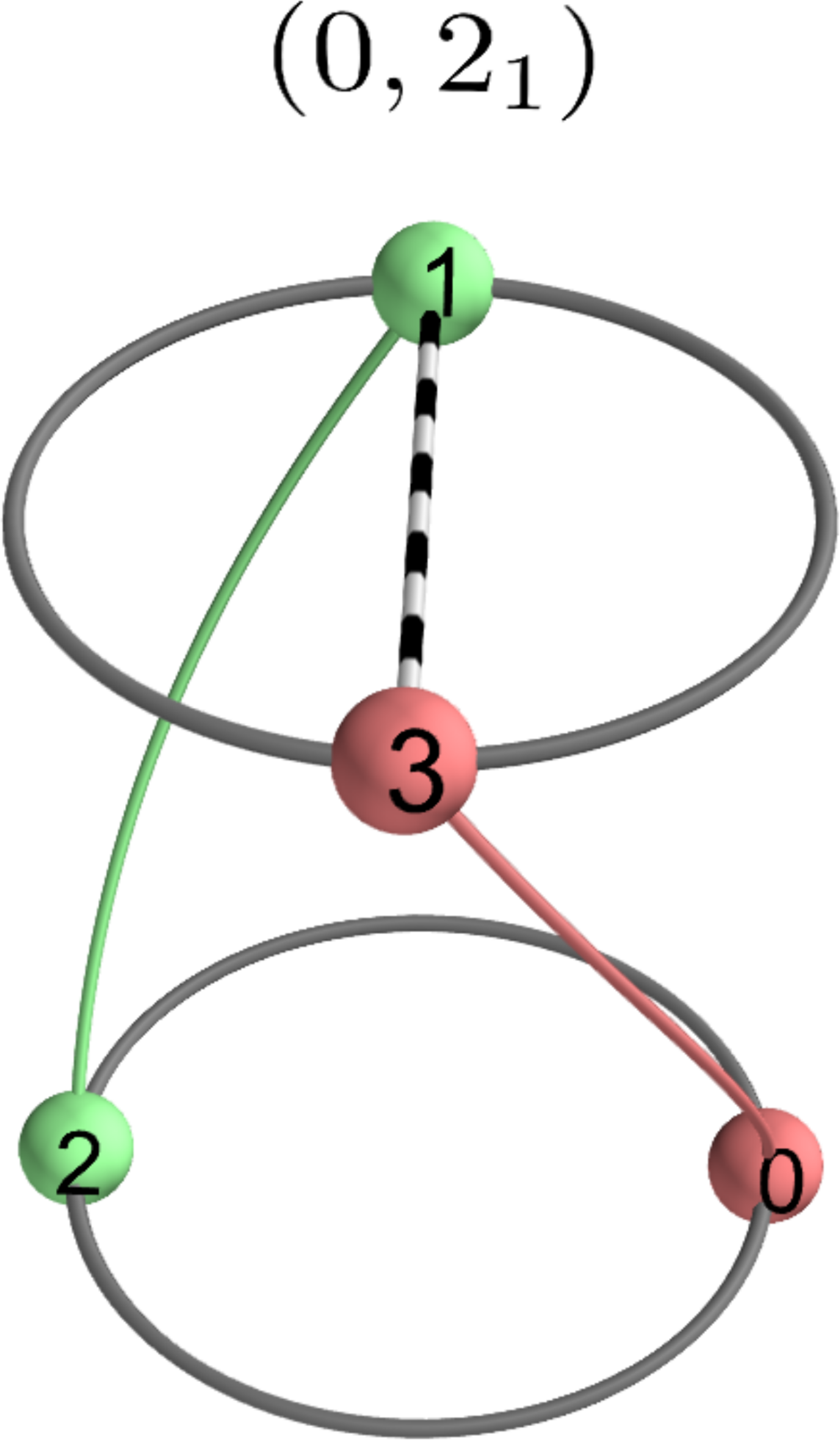}\quad
{\bf (b)} 
\includegraphics[width=4.4cm]{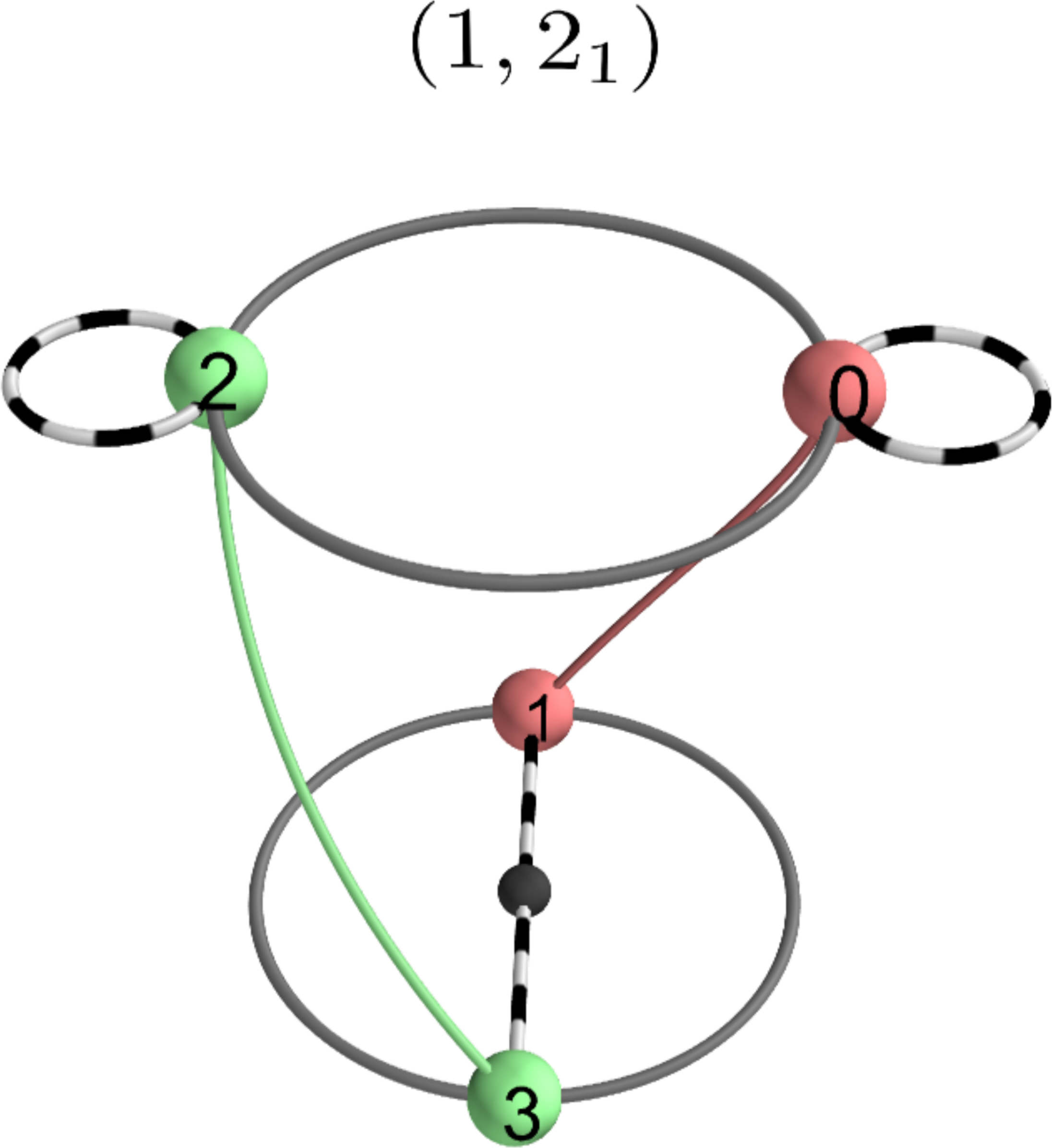}
\end{center}
\caption{\label{fig3} (Color online) Same as Fig.~\ref{fig2}, but for
  a $2_1$-invariant axis. Dashed-line connectors without a marker in
  the middle represent nonchiral degeneracies caused by band-gluing
  across the BZ face orthogonal to the symmetry axis.}
\end{figure}

Consider now the $(0,2_1)$ diagram shown in Fig.~\ref{fig3}(a).  The
symmetry labels at $\kz=0$ are $j=0,2$. Since $F=0$ and these labels
lie on the equator, the states are nondegenerate; this explains their
absence from Table~\ref{table:spinless}.  At $\kz=\pi$ the labels are
$j=1,3$, and they also do not appear in Table~\ref{table:spinless}
(where there are no entries with $n=2$) even though such states must
be pairwise degenerate according to the Wigner rules. The reason is
that the degeneracy is not an isolated Weyl node. Instead, the bands
remain glued together over the entire BZ
face~\cite{herring-pr37a,heine-book60}. A dashed line without a marker
is used to represent this nonchiral ``sticking of the bands.''

The $(1,2_1)$ diagram of Fig.~\ref{fig3}(b) shows that with spinful
$\T$ all bands pair up to form single Weyl nodes at $\kz=0$.  The
bands are again glued together over the entire BZ face at $k_z=\pi$,
but now the degenerate partners share the same label, $j=0$ or 2, on
the rotation axis.

Band gluing across a BZ face orthogonal to a $2_1$ axis occurs in
$\T$-invariant crystals because points on the BZ face are mapped onto
themselves by $\T*2_1$; the fact that this symmetry operation is
antiunitary and squares to $-1$ (for both $F=0$ and $F=1$) then forces
a Kramers degeneracy~\cite{sakurai-book94}.  Since the presence of
either $4_1$, $4_3$, $6_1$, $6_3$ or $6_5$ symmetry implies the
presence of $2_1$ symmetry, the band sticking occurs for all of them.
Indeed Eqs.~(\ref{eq:nnj}) and (\ref{eq:T-cond}) require $F'm$ to be
even when $n$ is even, and as a result,$f(q_+,q_-,0)$ vanishes when
$n$ is even, $m$ odd, and $\kz=\pi$. Together with
\eq{g-off-axis-trim}, this implies that when $n$ is even and $m$ is
odd the dispersion on the $k_z=\pi$ plane is described by
$H_{\rm eff}(q_x,q_y,\pi)=d(q_x,q_y,0)\mathbbm{1}$, confirming that
the bands remain glued together across the BZ face.  The combination
of~$\T$ with screw symmetries other than the ones listed above (i.e,
with $3_1$, $3_2$, $4_2$, $6_2$, or $6_4$) does not provoke
band-glueing and can stabilize Weyl nodes at $\kz=\pi$, as we already
saw for $3_1$.

The complete set of $(F,n_m)$ diagrams is given in
Appendix~\ref{app:all-diag}.  A few special cases of our systematic
classification have been noted in the recent literature. Double Weyl
nodes protected by $3_0$ and spinless $\T$ symmetry [see the $(0,3_0)$
diagram in Fig.~\ref{fig7}] were treated in
Ref.~\cite{chen-natcomm16}, and triple nodes protected by 6-fold
symmetry and spinful $\T$ symmetry are mentioned in
Ref.~\cite{chang-arxiv16}.

\section{Application to trigonal tellurium}
\label{sec:tellurium}

Elemental Te is a nonmagnetic semiconductor that crystalizes in two
enantiomorphic structures with space groups P$3_1$21 (No. 152,
right-handed) and P$3_2$21 (No. 154, left-handed).  The unit cell
contains three atoms disposed along a spiral chain, with the chains
arranged on a hexagonal net. The structure and its symmetries are
detailed in Refs.~\cite{asendorf-jcp57,nussbaum-procIRE62}, where it
can be seen that the spiral chains reduce the symmetry from hexagonal
to trigonal. In the following we pick right-handed Te and classify the
Weyl crossings along the trigonal axis $\Gamma$A in the hexagonal BZ
shown in Fig.~\ref{fig1}(a). (For left-handed Te the band structure is
identical, but the chiral charges flip sign.)

The valence-band maximum and conduction-band minimum of trigonal Te
occur close to the H point on the HK line. Without spin-orbit coupling
the conduction-band minimum is exactly at H, and with spin-orbit the
topmost valence band has a ``camelback'' shape, with a local minimum
at H surrounded by two maxima along
HK~\cite{natori-jpsj75,peng-prb14}.  States along the $\Gamma$A line
are far from the band edges and hence do not participate in the
low-energy physics.  We will study them with the sole purpose of
illustrating our classification scheme for Weyl nodes.

\subsection{Spinless bands}
\label{sec:Te-spinless}

\begin{figure}
\begin{center}
\includegraphics[width=1.0\columnwidth]{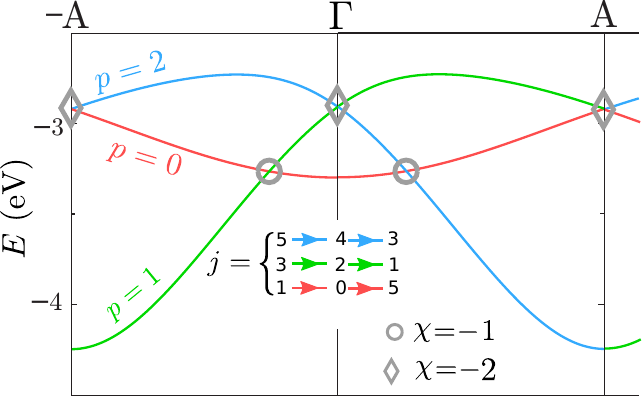}
\end{center}
\caption{\label{fig4} (Color online) A connected group of three
  valence bands in trigonal tellurium along the rotationally-invariant
  line $\Gamma$A, calculated without including spin-orbit coupling.
  Energies are measured from the valence-band maximum, and each color
  denotes a branch labeled by the integer $p$ in \eq{rot-eig}. For
  each branch, the values of the label $j$ in \eq{rot-eig-trim} at
  $-$A, $\Gamma$ and A (respectively, $\kz=-\pi,0$ and $\pi$) are also
  indicated.  Markers denote Weyl crossings with chiral
  charges~$\chi$.}
\end{figure}

We begin with a calculation that does not include spin-orbit coupling.
The bands split into ``elementary
representations''~\cite{michel-prb99} containing three bands each, and
in Fig.~\ref{fig4} we plot along $\Gamma{\rm A}$ the second-highest
valence-band complex.  As in Fig.~\ref{fig1}, each color denotes a
branch labeled by the integer $p$ in \eq{rot-eig}, with the branch
cuts chosen at $\kz=\pi$ mod~$2\pi$.  The labels were determined in
two ways: (i) by direct calculation starting from the wave functions,
and (ii) using $p=j(\Gamma)/2$ [\eq{rot-eig-trim}], after determining
the $j$ labels at $\Gamma$ and A as explained below.

It can be seen from \eq{rot-eig} that as $\kz$ changes by $2\pi$,
branch $p$ connects with branch $p-m$ mod~$n$, which in the present
case amounts to $p-1$ mod~3.  This is the monodromy phenomenon
described in Ref.~\cite{michel-prb99}, and it implies that the three
bands must be connected along the $\Gamma$A line in such a way that
one can travel continuously through all of them. The argument only
relies on screw symmetry and is silent on the nature and location of
the contact points, which also depend on the $\T$
symmetry~\cite{michel-prb99}.

Let us first classify the degeneracies at $\Gamma$ and A in
Fig.~\ref{fig4}, with the help of the $(0,3_1)$ diagram in
Fig. \ref{fig2}(a).  The band that is nondegenerate at $\Gamma$ has
$j=0$, and the two degenerate bands have $j=2$ and~4. The $j=4$ state
evolves to become the nondegenerate $j=3$ state at A, while $j=2$
evolves into $j=1$ to become degenerate with the $j=5$ state that
evolved from $j=0$.  The two degenerate pairs, one at $\Gamma$ and one
at A, form double Weyl nodes. By consulting Table~\ref{table:spinless}
we conclude that both have negative chirality [for the node at
$\Gamma$ (A), the higher-energy state at $\kz=0^+$ ($\kz=\pi^+$) has
label $j=2$ ($j=5$)].  As in Sec.~\ref{sec:cobalt} we have checked
these results by calculating the chiral charges explicitly from the
Berry curvature, and the same was done for the other cases discussed
below.

We have been assuming that the only symmetries present at $\Gamma$ and
A are $3_1$ and $\T$, when in fact those points are also left
invariant under 2-fold rotations along the $\Gamma$K and AH axes
respectively~\cite{asendorf-jcp57,nussbaum-procIRE62}. This does not
change our conclusions, because the presence of $2_0$ symmetry does
not lead to degeneracies at either $\Gamma$ or A, as can be seen from
the $(0,2_0)$ diagram in Fig. \ref{fig7} of Appendix
\ref{app:all-diag}. Hence, the degeneracies that do occur at those
symmetry points are correctly described by the $(0,3_1)$ diagram.

Let us now determine the chiral charges of the two non-TRIM crossings
in Fig.~\ref{fig4}; since they are related by a 2-fold rotation it is
sufficient to focus on one of them, e.g., that between the red and
blue branches on the right-hand side of the figure.  Recalling that
$\ket{u}$ and $\ket{v}$ are respectively the higher- and lower-energy
states after the crossing, we assign $p_u=0$ and $p_v=2$.
\equ{rot-ratio} then gives $\alpha_u/\alpha_v=e^{i2\pi/3}$, and
consulting Table~\ref{table:generic} we find $\chi=-1$, in agreement
with the calculated value.

In this particular example, it was possible to characterize all the
contact points on the symmetry axis without having to calculate
explicitly from the wave functions either their chiral charges or the
rotational symmetry labels of the crossing bands.  While the crossings
at $\Gamma$ and A are topologically required by the monodromy
argument, those at intermediate $\kz$ values can be eliminated by
changing the Hamiltonian without changing the
symmetry~\cite{michel-prb99}. This is discussed further in
Appendix~\ref{app:Te-widerange}.

\begin{figure}
\begin{center}
\includegraphics[width=1.0\columnwidth]{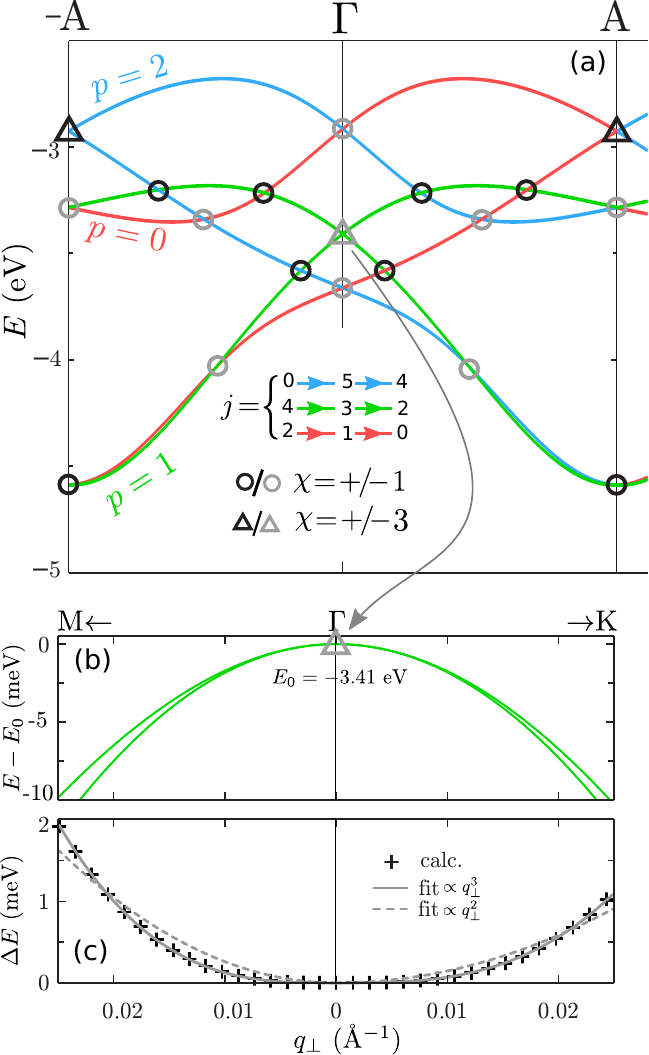}
\end{center}
\caption{\label{fig5} (Color online) (a) Same as Fig.~\ref{fig4} but
  with spin-orbit coupling included, resulting in a connected group of
  six bands. (b) In-plane dispersion near the triple Weyl point at
  $\Gamma$.  (c)~The corresponding splitting of the bands, together
  with its cubic and quadratic best fits. The band splittings along
  the $\Gamma$M and $\Gamma$K directions were fitted separately.}
\end{figure}

\subsection{Spinful bands}
\label{sec:Te-spinful}

Upon inclusion of spin-orbit coupling the 3-band complex of
Fig.~\ref{fig4} turns into the 6-band complex of
Fig.~\ref{fig5}(a). The branches are color-coded by the symmetry
labels $p=0,1,2$ in the same way as in Fig.~\ref{fig4}, with two
branches for each $p$.  Let us label the bands, ordered in energy at
each $\kk$, from one to six.  If we start at $-$A on the sixth band
and follow the topmost $p=2$ branch from $\kz=-\pi$ to $\kz=\pi$, it
connects with the fourth band on a $p=1$ branch. After one more
monodromy cycle that branch connects with the the second band at $-$A
on a $p=0$ branch, which connects back with the original $p=2$ branch
on the sixth band after a third cycle. So far we have only covered
half of the band complex; in order to span the other half (comprising
three more branches with $p=0,1,2$), we can carry out a new sequence
of monodromy cycles starting from the fifth band at $-$A.  Each band
is split in half between the two groups of three branches, and the two
groups are connected to one another by the continuity in $\kk$ of
energy bands.

As in the spinless case, the qualitative features of the degeneracies
at $\Gamma$ and A can be inferred by simply inspecting the band
structure and referring to the corresponding diagram in
Fig.~\ref{fig2}(b).  According to that diagram, Weyl nodes are
unavoidable at both points; since all allowed~$j$ labels occur the
same number of times within the complex, at each TRIM two of the nodes
must be single Weyl nodes and one a triple node. Note also that states
forming the triple node at one TRIM must hook up with states belonging
to different single nodes at the other. It is then sufficient, in
order to label all six bands, to identify one of the triple nodes
(e.g., by examining the in-plane band splittings and selecting the one
that is nonlinear). Suppose we have established that the triple node
at $\Gamma$ is the middle one in energy, as indicated in the figure.
Since it connects with the two lower-energy nodes at~A, these must be
single nodes; as expected, the remaining (triple) node at~A connects
with the two single nodes at $\Gamma$.  We can now assign all the $j$
labels at~$\Gamma$ and~A, and Table~\ref{table:spinful} then gives the
chiralities of all four single nodes at those two points (but not of
the two triple nodes, whose chiralities are not fixed by the symmetry
labels and had to be determined from the Berry flux).

Figures~\ref{fig5}(b) and~\ref{fig5}(c) show the in-plane dispersion
and splitting of the bands around the triple Weyl node at $\Gamma$.
In this case the splitting is cubic as predicted in
Sec.~\ref{sec:timerev-symm-formal}, not quadratic as for triple nodes
occurring at generic points along a 6-fold axis [e.g.,
Fig.~\ref{fig1}(e)].

Finally, consulting Table~\ref{table:generic} we find the chiralities
of the five band crossings between~$\Gamma$ and~A in Fig.~\ref{fig5},
which must be single Weyl nodes since the axis has 3-fold symmetry.
Amusingly, the rules for determining the chiralities in this case are
those of the game
``Rock--paper--scissors''\cite{wikipedia-rock-paper-scissors}; a red
branch is ``rock,'' blue is ``paper,'' and green is ``scissors.''  If
the higher-energy state after the crossing is the winner then
$\chi=+1$, otherwise $\chi=-1$; since branches of the same color do
not cross, there can be no tie.

In Appendix~\ref{app:Te-widerange}, we repeat the above analysis for
two more six-band complexes in the spinor band structure of Te, and
before closing this section we mention that Weyl nodes in Te were
studied in Refs.~\cite{hirayama-prl15,nakayama-prb17}. The main focus
of both works was on the nodes occurring close to the valence-band
maximum and conduction-band minimum near the H point. The existence of
triple Weyl nodes at $\Gamma$ and A is not mentioned in either work,
where all the reported Weyl points have $\vert\chi\vert=1$.

\section{Effect of a  $\T$-breaking perturbation} 
\label{sec:trbreak} 

\begin{figure}
\includegraphics[width=\columnwidth]{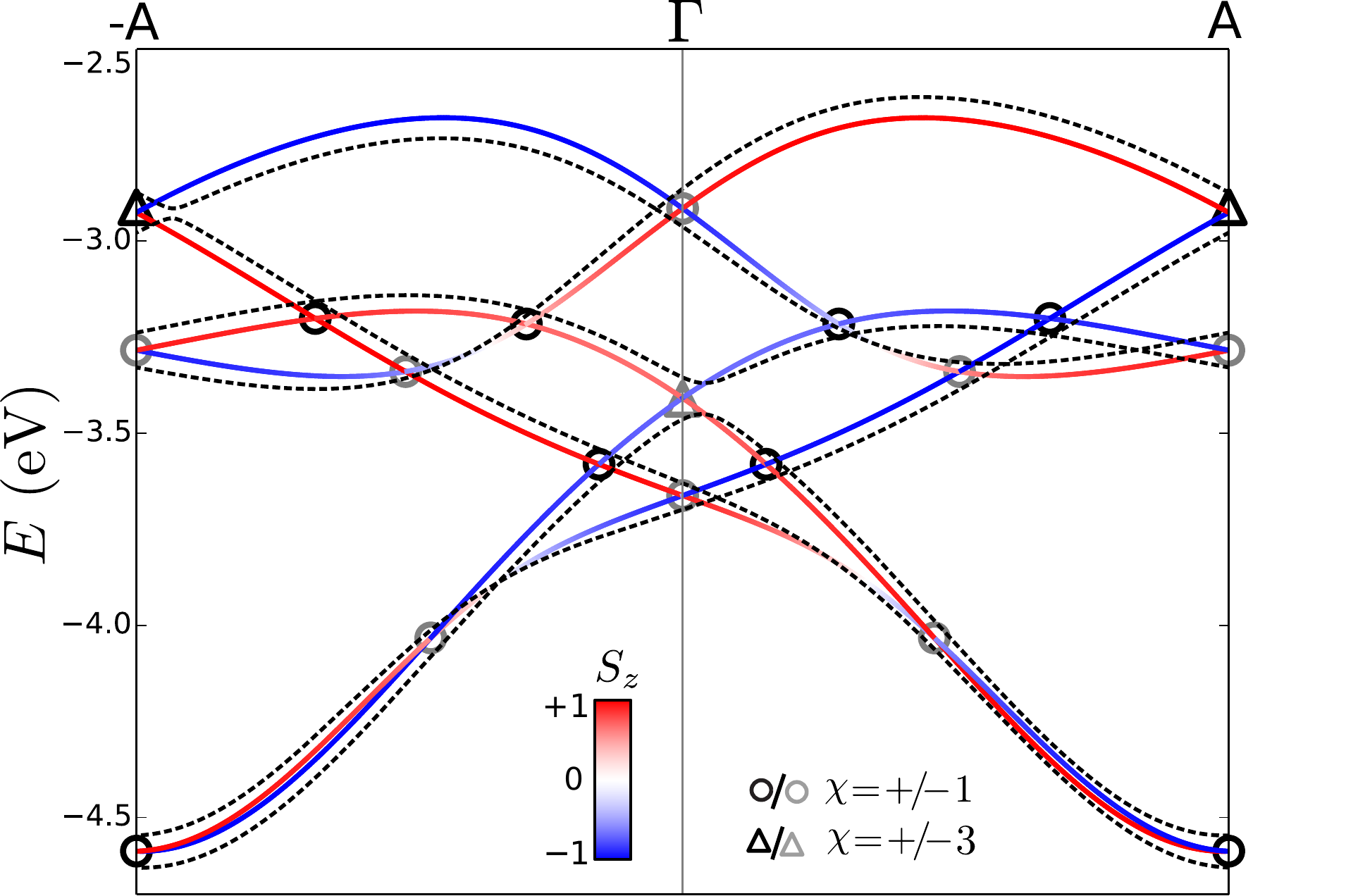}
\caption{(Color online) Solid lines: ground-state band structure of Te
  color-coded by the expectation value of the spin (in units of
  $\hbar/2$) projected along the trigonal axis $\zhat$.  Dashed lines:
  the band structure in the presence of a macroscopic magnetization
  amounting to a magnetic moment per atom of
  $\mu_z=0.01\mu_\mathrm{B}$. Symbols denote Weyl nodes in the ground
  state with ${\bm \mu}=0$, and energies are measured from the
  ground-state valence-band maximum. \label{fig6} }
\end{figure}

In this section we continue to focus on acentric crystals with a
hexagonal BZ and a nonmagnetic ground state, and study how spinful
Weyl nodes on the $\Gamma$A line are effected by a perturbation that
breaks $\T$ symmetry but preserves the rotational symmetry.  Because
only 6-fold (not 3-fold) symmetry is able to protect triple Weyl nodes
in the absence of $\T$ symmetry, the effect of such a perturbation on
a triple node pinned to either $\Gamma$ or A is different depending on
whether $\Gamma$A is a 3-fold or a 6-fold axis. In order to illustrate
the two types of behaviors, we consider below the cases of trigonal Te
and hexagonal NbSi$_2$.

\subsection{Trigonal Tellurium}

We start by recalculating the spinor band structure of Te in the
presence of a weak local Zeeman field directed along the trigonal axis
$\zhat$, which induces a finite magnetic moment ${\bm \mu}=\mu_z\zhat$
on each Te atom. Technically, this is achieved by adding a ``penalty
term'' of the form
$\lambda\sum_{i=1}^3\vert {\bm \mu}-{\bm \mu}_i\vert^2$ to the density
functional during the self-consistent loop. Here, ${\bm \mu}_i$ is the
self-consistent magnetic moment on the $i$-th atom in the unit cell,
and $\lambda>0$ is an adjustable parameter.

To lowest order, the contribution of the induced magnetization to the
Hamiltonian takes the form
\begin{equation}
\label{eq:dHM}
\Delta H\propto M_z \tau_z
\end{equation}
in the basis of the unperturbated eigenstates. Here, $M_z$ is the
magnetization and $\tau_z$ is the Pauli matrix describing the spin
degree of freedom, not to be confused with the pseudospin matrix
$\sigma_z$ in \eq{ham-eff}.  Since our discussion will be qualitative,
we are content to leave \eq{dHM} expressed as a proportionality.

Figure~\ref{fig6} shows how the band structure on the $\Gamma$A line
changes in the presence of a small positive $M_z$.  The ``generic''
Weyl nodes away from $\Gamma$ and A, which are pinned to the symmetry
axis by the $3_1$ symmetry, remain on the axis because that symmetry
is preserved with ${\bm M}\parallel\zhat$.  However, they move in
energy and wavevector, in a way that depends on the spin projections
of the two crossing states.  This behavior can be understood from the
effective two-band Hamiltonian near a generic Weyl node of charge
$\chi=\mp 1$, which reads
\begin{equation}
\label{eq:ham-zeeman}
H_{\rm eff}(q_+,q_-,q_z)=
\left(
  \begin{array}{cc}
    \alpha q_z+M_z s_u& 2aq_\pm \\ 2a^*q_\mp & -\alpha q_z+M_z s_v 
  \end{array}  
\right)
\end{equation}
with the zero of energy placed at the unperturbed crossing.  Here
$s_u=\me{u}{\tau_z}{u}$ and $s_v=\me{v}{\tau_z}{v}$ are the spin
projections of the upper- and lower-energy states on the right-hand
side of the crossing. Note that the perturbation~(\ref{eq:dHM}) does
not introduce off-diagonal terms in the effective Hamiltonian: since
$p_u\neq p_v$ for single Weyl nodes, $\me{u}{\tau_z}{v}$ vanishes
according to Appendix~\ref{app:level-repulsion}. For $M_z=0$ and
$q_z=0$ \eq{ham-zeeman} reduces to the Hamiltonian given in the second
row of Table~\ref{table:generic}; for $M_z=0$ and $q_+=q_-=0$ it
reduces to \eq{ham-eff-parallel}, with the positive coefficient
$B_{001}$ written here as~$\alpha$.

Assuming the crossing states are the eigenstates of the spin
projection $S_z$ (implying $|s_u|=|s_v|=1$), we can derive from
\eq{ham-zeeman} with $q_+=q_-=0$ the following rules for how generic
Weyl nodes on the $\Gamma$A line shift under a small positive $M_z$:
\begin{center}
\begin{tabular}{c|c|cl}
$\mathrm{sgn}(s_u)$&$\mathrm{sgn}(s_v)$&\hspace*{0.5cm}& direction of shift\\
\hline
$+$&$+$& &to higher energy\\
$-$&$-$& &to lower energy\\
$+$&$-$& &to lower $\kz$\\
$-$&$+$& &to higher $\kz$
\end{tabular}
\end{center}
Even though the crossing states are generally not eigenstates of
$S_z$, these simple rules describe fairly well the shifts of most of
generic nodes in Fig.~\ref{fig6}.

Let us turn now to the Weyl nodes pinned to $\Gamma$ and A by $\T$
symmetry. In this case the unpreturbed crossing states are
time-reversal partners, so that $s_v=-s_u$.  Consider first the nodes
with $\vert\chi\vert=1$.  When $\T$ is broken by $M_z$, those nodes
move away from the TRIM but remain on the $\Gamma$A axis by virtue of
the unbroken $3_1$ symmetry, and their motion along the axis obeys the
same rules discussed above for the generic single Weyl nodes. Note
that the positive node at A and $\sim -4.5$~eV moves to the left and
anihilates with the negative node at $\sim -4.0$~eV and halfway
between $\Gamma$ and A, which moves to the right.

The most striking effect of the axial Zeeman field is on the triple
Weyl nodes at $\Gamma$ and A.  In Fig.~\ref{fig6}, gaps open up near
those nodes when $M_z\not=0$.  That happens because $\T$ symmetry not
only pins those nodes to the TRIM, but is essential for their
existence on a 3-fold axis (recall that without $\T$ symmetry, triple
nodes can only occur on 6-fold axes).  Once the $\T$ symmetry is
broken, each triple node splits into three single nodes; these are not
visible in Fig.~\ref{fig6} because they are located off the $\Gamma$A
axis on $\Gamma$MLA planes (see below).

In order to understand the splitting pattern, let us write the
effective Hamiltonian for a triple node including the perturbing
term~(\ref{eq:dHM}).  Taking the unperturbed terms with $q_z=0$ from
the third row in Table~\ref{table:spinful} and the unperturbed terms
with $q_+=q_-=0$ from \eq{f-z} we find, setting $A_{001}=\beta$ and
using $s_v=-s_u$,
\begin{multline}
\label{eq:ham_wp3}
H_{\rm eff}(q_+,q_-,q_z)=(\alpha q_z+cq_+^3+c^*q_-^3+s_u M_z)\sigma_z +{}\\
{}\left[(aq_+^3+bq_-^3+\beta q_z +\gamma M_z)\sigma_+  
+\mathrm{H.c.}\right],
\end{multline}
where $\gamma=\frac{1}{2}\me{u}{\tau_z}{v}$ and ``H.c.''  stands for
Hermitian conjugate. Without additional symmetries all coefficients
here are generally complex, except for $\alpha$ which is real and
positive.  The condition for a degeneracy to occur is that the
prefactors of the $\sigma_z$ and $\sigma_\pm$ matrices should vanish
simultaneously.  Thus
\beq
\label{eq:shift-qz}
q_z=-\frac{1}{\alpha}(s_u M_z+cq_+^3+c^*q_-^3),
\eeq
which inserted into the prefactor of $\sigma_+$ gives
\begin{equation}
\wt aq_+^3+\wt bq_-^3+\wt\gamma M_z=0,
\end{equation}
where $\wt a =a -c\beta/\alpha$, $\wt b =b -c^*\beta/\alpha$ and
$\wt \gamma = \gamma - \beta s_u/\alpha$.  Writing $q_\pm$ as
$q_\perp e^{\pm i\phi}$ with $q_\perp>0$ leads to
\begin{equation}
 \label{eq:phiq}
q_\perp^3(\wt a e^{i3\phi}+\wt be^{-i3\phi})+\wt \gamma M_z=0.
\end{equation}
The condition
$\mathrm{arg}(\wt a e^{i3\phi} +\wt b
e^{-i3\phi})=\mathrm{arg}(-\wt\gamma M_z)$
for the phase has three roots separated by $2\pi/3$. This means that
the in-plane splitting pattern of a triple node respects 3-fold
symmetry, as expected.  For the magnitude of the splitting we get
\begin{equation}
  q_\perp^3=
  \left|
    \frac{\wt \gamma M_z}{\wt a e^{i3\phi}+\wt b e^{-i3\phi}}
  \right|.
\end{equation}
Thus the in-plane splitting increases as $\vert M_z\vert^{1/3}$, and
from \eq{shift-qz} we conclude that the vertical shift of the three
split nodes is instead linear in $\vert M_z\vert$.  However, unlike
for a single Weyl node, the direction of the on-axis shift of a triple
node in not uniquely defined by the spin projections of the crossing
states, but also depends on microscopic parameters.

The previous analysis correctly predicts a 3-fold symmetric in-plane
splitting of a $\T$-invariant triple node on a 3-fold axis, but it
does not determine its absolute orientation. In order to do so, it is
necessary to take into consideration an additional symmetry of
trigonal Te, namely the two-fold rotation.  In
Appendix~\ref{app:splitdir} we show that this symmetry pins the three
split Weyl nodes to the $\Gamma$MLA planes, as confirmed by the
first-principles calculations.

\subsection{Hexagonal NbSi$_2$}
\label{sec:NbSi2}

We take as our last example the transition-metal silicide NbSi$_2$, a
nonmagnetic metal that crystallizes in a noncentrosymmetric hexagonal
structure with three formula units per cell~\cite{antonov-prb96}. The
structure is enantiomorphic, and we pick the right-handed variety with
space group P6$_2$22 (No.~180). Due to the combination of a six-fold
rotational symmetry and time-reversal invariance, the energy bands of
NbSi$_2$ contain single and triple nodes at the points $\Gamma$ and A,
as well single, double and triple nodes at generic points along the
$\Gamma$A line. In the following we focus on the Weyl nodes pinned to
the $\Gamma$ point, and study numerically how they are affected by a
Zeeman field directed along the hexagonal axis.

\begin{figure}
\includegraphics[width=\columnwidth]{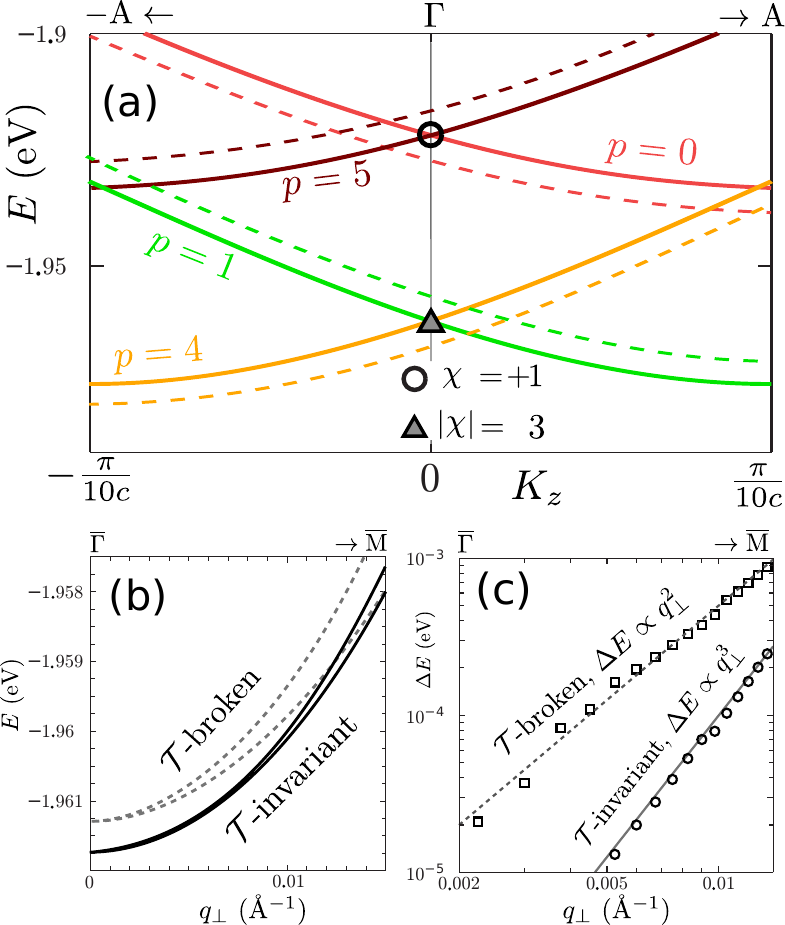}
\caption{\label{fig:NbSi2} (Color online) (a) Spinor band structure of
  NbSi$_2$ in the vicinity of a pair of single and triple Weyl nodes
  at~$\Gamma$. Solid lines: Band structure in the nonmagnetic ground
  state, color-coded in a similar way as Figs.~\ref{fig4}
  and~\ref{fig5}(a). Dashed lines: Band structure with $\T$ symmetry
  broken by an induced magnetic moment of 0.1~$\mu_{\rm B}$ per unit
  cell.  (b) shows the in-plane dispersion near the triple node,
  with (dashed lines) and without (solid lines) the induced magnetic
  moment. The corresponding in-plane band splittings away from the
  triple node are shown in panel~(c) on a double logarithmic scale,
  together with their quadratic and cubic fits.}
\end{figure}

In Fig.~\ref{fig:NbSi2}(a), we plot along $\Gamma$A four spinor bands
below the Fermi level in the vicinity of~$\Gamma$. The unperturbed
bands are drawn as solid lines, with each color denoting a branch
labeled by the integer $p$ in \eq{rot-eig}, which was determined from
the calculated Bloch states. The states at $\Gamma$ are more
conveniently labeled using \eq{rot-eig-trim}: $j(\Gamma)=2p+1$. In
accordance with the $(1,6_2)$ diagram in Fig.~\ref{fig8}, the two
upper bands with $j=1$ and $11$ form a single Weyl node, while the two
lower bands with $j=3$ and $9$ form a triple node.
Table~\ref{table:spinful} tells us that the single node has positive
chirality, and that the chirality of the triple node is not fixed by
the rotational symmetry of the crossing states (in principle it could
be determined from the Berry flux, but we have not done so).

The bands perturbed by the Zeeman field are plotted in
Fig.~\ref{fig:NbSi2}(a) as dashed lines.  While the effect on the
single Weyl node at $\Gamma$ is the same as in Fig.~\ref{fig6} for
trigonal Te (the node shifts away from the TRIM along the axis), the
effect on the triple node is different.  Instead of splitting into
three off-axis single nodes, the triple node simply moves away from
the TRIM along the axis.  This is due to the fact that contrary to the
3-fold symmetry of Te, the 6-fold symmetry of NbSi$_2$ is able to
stabilize a triple Weyl node without the assistance of $\T$ symmetry,
whose only effect is to pin the triple node to a TRIM and to modify
its in-plane dispersion. Figures~\ref{fig:NbSi2}(b,c) demonstrate that
once $\T$ is broken the in-plane band splitting changes from cubic to
quadratic, in accordance with Tables~\ref{table:generic}
and~\ref{table:spinful}.

\section{Conclusions}
\label{sec:conclusions}

We have carried out a systematic classification of the types of
degeneracies that can occur on symmetry lines in the BZ of a 3D
crystal with pure rotational or screw symmetry $n_m$, assuming broken
$\P*\T$ symmetry and in the absence of other crystallographic
symmetries.  We first presented the classification for the generic
case, and then specialized to $\T$-invariant, $\P$-broken crystals,
treating both spinless and spinful $\T$. At generic points along a
symmetry axis the degeneracies are Weyl nodes. At $\T$-invariant
points on an axis they can be either Weyl nodes or, when $n$ is even
and $m$ is odd, gluing of pairs of bands extending over the entire
perpendicular BZ face.

It was known from previous work~\cite{fang-prl12} that the presence of
either 4-fold or 6-fold rotational symmetry without $\T$ symmetry can
stabilize Weyl nodes with a chiral charge of magnitude larger than
one: $\vert\chi\vert=2$ when $n=4$ or $6$, and also $\vert\chi\vert=3$
when $n=6$. In this work we have found a new type of triple Weyl node
that
is stabilized by spinful $\T$ symmetry in combination with either
$3_m$ or $6_m$ symmetry; for $6_1$, $6_3$, and $6_5$ nodes of this
type can only occur at~$\Gamma$, while in all other cases they can
also occur at the A point.  The two types are qualitatively different;
generic triple nodes protected by 6-fold symmetry alone away from the
TRIM are {\it dressed}, with the in-plane cubic chiral dispersion
masked by a nonchiral quadratic dispersion; thus, at leading order the
band splitting is indistinguishable from that of a double Weyl node.
Instead, $\T$-invariant triple nodes at $\Gamma$ or A on a 3-fold or a
6-fold axis are {\it naked}: the leading-order in-plane splitting is
cubic. It follows that there is no one-to-one correspondence between
the chiral charge of a Weyl node on a symmetry axis and the
leading-order in-plane splitting of the bands. Ferromagnetic hcp Co
and trigonal Te are examples of materials possessing dressed and naked
triple Weyl nodes, respectively, while both types are present in
hexagonal NbSi$_2$.

In summary, we have shown that composite Weyl nodes are both more
common and more diverse than previously thought, and that in some
cases they are unavoidable. We hope that these findings will stimulate
the search for new classes of Weyl semimetals where $\T$-invariant
Weyl nodes occur on symmetry axes and close to the Fermi level.

\begin{acknowledgments}

  S.S.T and I.S. acknowledge support from Grant No.~FIS2016-77188-P
  from the Spanish Ministerio de Econom\'ia y Competitividad, Grant
  No. CIG-303602 from the European Commission, and from Elkartek Grant
  No. KK-2016/00025. D.V. acknowledges support from NSF Grant
  DMR-1408838 and thanks J. Bonini for useful discussions.  Computing
  resources were provided by the Donostia International Physics
  Centre.

\end{acknowledgments}

\appendix

\section{Level repulsion versus level crossing at generic points on a
  rotation axis}
\label{app:level-repulsion}

Let $\ket{u}$ and $\ket{v}$ be two distinct eigenstates of the screw
symmetry operator $C_{n_m}$ at a generic point $\kz$ along the
rotation axis of \eq{k-axis}, and let $\alpha_{p_u}$ and
$\alpha_{p_v}$ be the corresponding eigenvalues given by
\eq{rot-eig}. Suppose the two states are also degenerate eigenstates
of the Hamiltonian $H$, which by assumption commutes with $C_{n_m}$.
Now add a small perturbation $\Delta H$ that respects the $C_{n_m}$
symmetry. We want to know the conditions under which $\Delta H$ can
couple the two states and open a gap. If the crossing can be removed
in this way then it means that the original Hamiltonian $H$ was
fine-tuned.

The coupling matrix element is
\bea
\label{eq:coupling}
\me{v}{\Delta H}{u}&=&\me{v}{C_{n_m}^{-1}
\left( C_{n_m}\Delta H C_{n_m}^{-1}\right)C_{n_m}}{u}\nn
&=&e^{i2\pi(p_u-p_v)/n}\me{v}{\Delta H}{u},
\eea
where we used $[\Delta H,C_{n_m}]=0$. There are two cases:
\begin{enumerate}

\item If $p_u\not=p_v$~mod~$n$ \eq{coupling} can only be satisfied
  with $\me{v}{\Delta H}{u}=0$, which means that the perturbation does
  not open a gap: The crossing is robust against symmetry-preserving
  perturbations.

\item If $p_u=p_v$ mod~$n$ then $\me{v}{\Delta H}{u}$ can be nonzero,
  and $\Delta H$ will generically split the degeneracy.

\end{enumerate}

\section{\label{app:methods} Details of the numerical calculations }

\subsection{Ground state calculations}

The electronic structure calculations of Secs.~\ref{sec:cobalt}
and~\ref{sec:tellurium} were carried out within the framework of
density-functional theory, as implemented in the {\tt VASP} code
package~\cite{VASP1,VASP2}. This code uses a plane-wave basis set to
expand the valence wave functions, and the projector-augmented wave
method to describe the core-valence interaction~\cite{PAW1,PAW2}.
Except for Sec.~\ref{sec:Te-spinless}, the calculations reported in
this work include spin-orbit coupling in the core-valence interaction.

Fully-relativistic total energy calculations for hcp Co in its
ferromagnetic ground state were carried out using the experimental
lattice parameters $a=2.506$~\AA~and $c=4.067$~\AA~\cite{Masumoto}.
Exchange and correlation effects were treated using the Perdew, Burke,
and Ernzerhof generalized-gradient approximation (GGA-PBE)~\cite{PBE}.

For the calculations on trigonal Te we used the experimental
parameters $a=4.458$~\AA~and $c=5.925$~\AA~\cite{Bouad2003189}, and a
relaxed value of $u=0.274$ for the dimensionless helix parameter,
which differs slightly from the experimental value of
$u=0.255$~\cite{Bouad2003189} (this parameter is defined as $u=r/a$,
with $r$ the radius of the helix~\cite{peng-prb14}).  Both the
generalized-gradient approximation and the local-density approximation
incorrectly predict a semi-metallic rather than semiconducting ground
state for this material, due to a closing of the gap at the H
point. Although this issue does not greatly affect our study of
degeneracies along the $\Gamma$A line, we have opted to correct it by
using instead the so-called HSE06 hybrid functional~\cite{HSE06}.  In
this way we obtained an energy gap of 0.312~eV at~H from a
fully-relativistic calculation, in good agreement with both the
calculated value of 0.314~eV obtained using the GW
method~\cite{hirayama-prl15} and the experimental value of
0.323~eV~\cite{Anzin1977}.

The band structure calculations for hexagonal NbSi$_2$ in the
nonmagnetic ground state were carried out using the GGA-PBE
approximation and the experimental lattice parameters
$a=4.819$~\AA~and $c=6.592$~\AA~\cite{Kubiak}.  In order to study the
effect of a $\T$-breaking perturbation, we added a non-selfconsistent
magnetization of 0.1$\mu_{\rm B}$ per unit cell.

\subsection{Post-processing using a Wannier-function basis}

\begin{figure}
\begin{center}
\includegraphics[height=14.5cm]{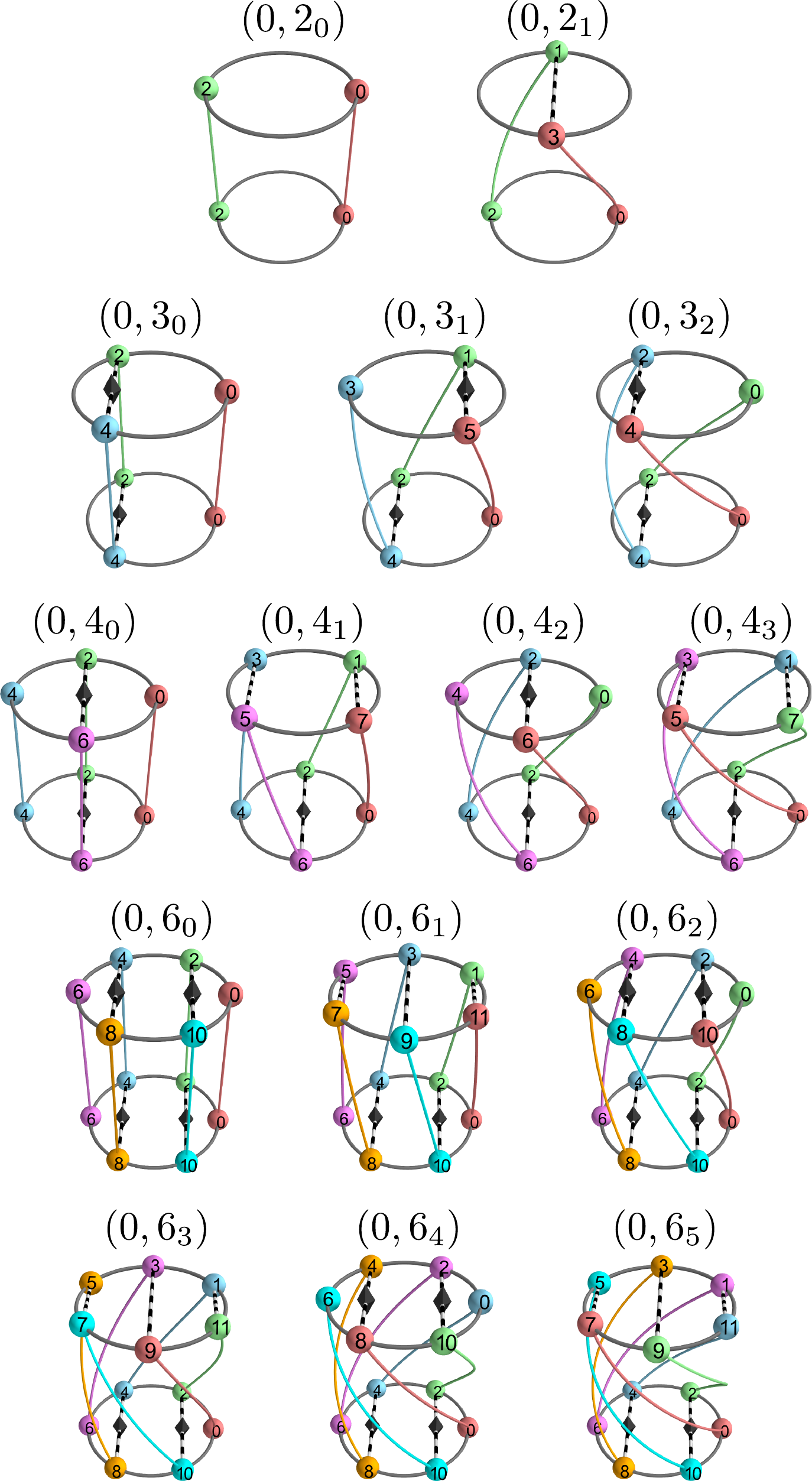}
\end{center}
\caption{\label{fig7} (Color online) Schematic representation of the
  types of degeneracies protected by $\T$ and $n_m$ symmetry at the
  TRIM $\kz=0$ and $\pi$ along an $n_m$-invariant axis $(0,0,\kz)$ in
  the BZ. Each diagram is labeled by $(F,n_m)$, and here we show the
  diagrams for spinless $\T$ ($F=0$) and all $n_m$ symmetries
  compatible with lattice periodicity. For a detailed explanation, see
  the captions of Figs.~\ref{fig2} and~\ref{fig3}.}
\end{figure}

In order to interpolate the energy bands and calculate the chiral
charges of the Weyl nodes, we use the formalism of maximally-localized
Wannier functions~\cite{PhysRevB.56.12847,PhysRevB.65.035109} as
implemented in the {\tt Wannier90} code
package~\cite{Mostofi2008,Mostofi2014}.

Exploring the band structure of hcp Co over the BZ we find that the
bands shown in Fig.~\ref{fig1} cross with higher-lying
bands~\cite{McMullan92}, and thus we use the disentanglement
procedure~\cite{PhysRevB.65.035109} to construct the Wannier
functions. The trial orbitals are chosen to be $sp^3$ hybrids and
atom-centered $d$ orbitals, totaling nine Wannier functions per atom
and spin channel. The outer energy window~\cite{PhysRevB.65.035109}
spans the range from $-20$~eV to $+70$~eV relative to the Fermi level,
which covers all $4s$, $4p$, and $3d$ states present in the
pseudopotential calculation, while the frozen energy
window~\cite{PhysRevB.65.035109} goes from $-20$ to $+7$~eV.
 
The $5p$ bands of trigonal Te are well separated from the lower $5s$
states, and they cross with higher-lying sates only in a small region
of the BZ. The outer energy window goes from $-8$ to $+5$~eV relative
to the valence-band maximum, the inner frozen window from $-8$~eV to
$+2.5$~eV, and we use atom-centered $p$-type trial orbitals for the
initial projections.  The resulting Wannier functions are similar to
those obtained in Ref.~\cite{hirayama-prl15} for the same material.

The chiral charges are calculated from the quantized Berry-curvature
flux through small surfaces enclosing the individual Weyl
nodes~\cite{gosalbez-prb15},
\begin{subequations}
\begin{align}
\chi_{l\alpha}&=\frac{1}{2\pi}\oint_S dS\,\hat{\bn}\cdot\bO_l(\bk)\\
\bO_l(\bk)&=i \langle \bnab_\bk u_{l\bk}|\times|\bnab_\bk u_{l\bk}\rangle.
\end{align}
\end{subequations}
Here, $\hat{\bn}$ is the unit surface normal pointing outwards, and
$l$ is the lower of the two bands that cross. In practice the Berry
curvature $\bO_l(\bk)$ is evaluated on a dense grid of $k$ points by
Wannier interpolation~\cite{wang-prb06}, and we choose the closed
surface $S$ to be a parallelepiped with the $\alpha$-th Weyl node
between bands $l$ and $l+1$ at the center~\cite{gosalbez-prb15}.

\begin{figure}
\begin{center}
\includegraphics[height=14.5cm]{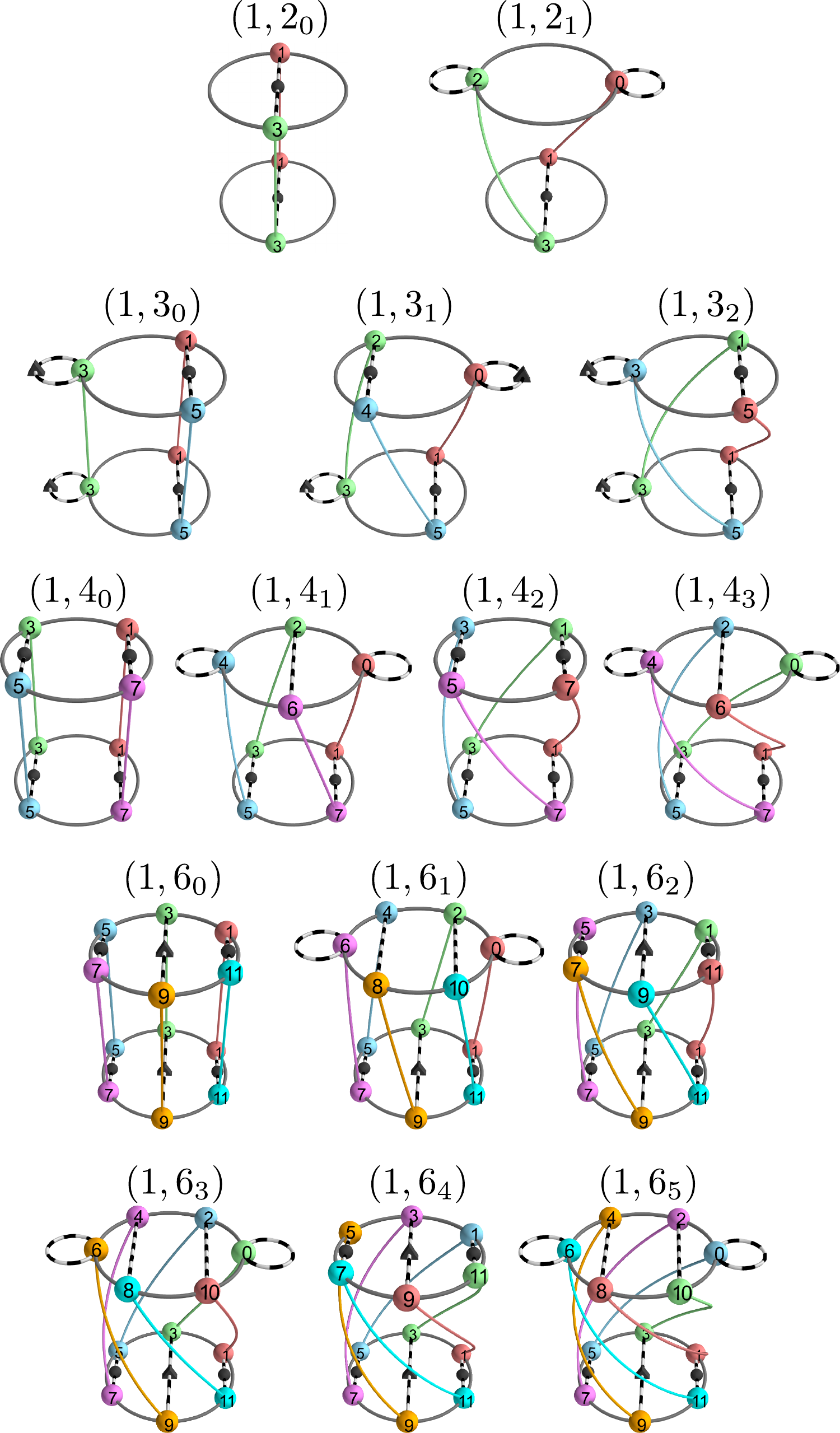}
\end{center}
\caption{\label{fig8} (Color online) Same as
  Fig.~\ref{fig7}, but for spinful $\T$ ($F=1$).}
\end{figure}

\section{Diagrams describing the types of degeneracies at TRIM on a
  rotation axis}
\label{app:all-diag}

We present here the complete set of $(F,n_m)$ diagrams introduced in
Sec.~\ref{sec:timerev-symm-diagr} to describe the degeneracies
occurring at TRIM on a rotation axis, assuming no other symmetries are
present.  The allowed degeneracies are Weyl points with chiral charges
of magnitude $\vert\chi\vert=1,2$ or 3, and band gluing extending
over the entire BZ face orthogonal to the symmetry axis.
Figure~\ref{fig7} contains the diagrams for spinless $\T$ ($F=0$), and
Fig.~\ref{fig8} contains the diagrams for spinful $\T$ ($F=1$).

The magnitude of the chiral charge has a simple interpretation in
terms of these diagrams: When $F=0$ ($F=1$), $\vert\chi\vert$ is equal
to the smallest nonzero even (odd) number of hops around the unit
circle needed to travel between spheres connected by a dashed line or
loop with a marker.  It follows that $\vert\chi\vert$ must be even
when $F=0$, and odd when $F=1$. For $n=2,3,4,6$, the only possible
values are $\vert\chi\vert =2$ for $F=0$ and $\vert\chi\vert =1,3$ for
$F=1$, with $\vert\chi\vert=3$ requiring $n=3$ or~$6$. These
conclusions are in agreement with Tables~\ref{table:spinless}
and~\ref{table:spinful}.

\section{Valence and low-lying conductions bands of tellurium }
\label{app:Te-widerange}

\begin{figure}
\includegraphics[width=\columnwidth]{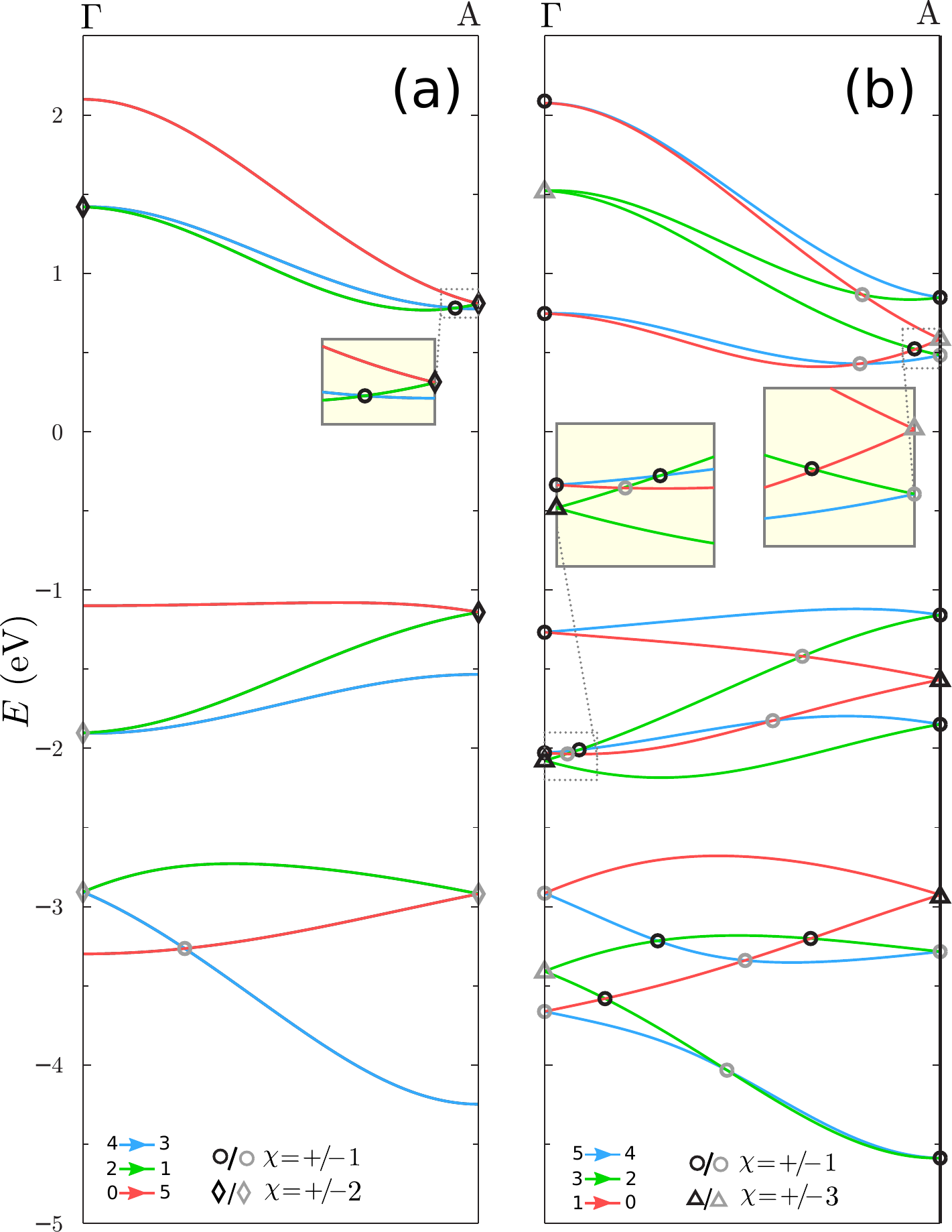}
\caption{(Color online) Three complexes (two valence and one
  conduction) in the band structure of Te, plotted along the 3-fold
  axis $\Gamma$A with energies measured from the valence-band
  maximum. The bands in~(a) and~(b) were calculated without and with
  spin-orbit coupling, respectively.  The meaning of the colors and
  markers is the same as in Figs.~\ref{fig4} and \ref{fig5}(a), which
  focus on the valence complex shown here at the bottom. \label{fig9}}
\end{figure}

We saw in Sec.~\ref{sec:Te-spinless} that the spinless band structure
of Te consists of 3-band complexes with topologically-required contact
points along the 3-fold axis $\Gamma$A. In addition to the double Weyl
nodes pinned to $\Gamma$ and A by $\T$~symmetry, there was an
additional Weyl crossing between $\Gamma$ and A in the complex of
Fig.~\ref{fig4}. This degeneracy is accidental, and can be eliminated
by changing the Hamiltonian without changing the
symmetry~\cite{michel-prb99}. This is corroborated by
Fig.~\ref{fig9}(a), where two more band complexes are shown. Both have
double Weyl nodes at $\Gamma$ and A, but the topmost valence complex
contains no accidental crossings in between. In this scenario with
minimal connectivity, the double nodes at $\Gamma$ and A must have
opposite chiralities, as can be seen from Table~\ref{table:spinless}.

Figure~\ref{fig9}(b) shows the spinful bands for the same three-band
complexes. The ordering in energy of the single and triple nodes at
$\Gamma$ and/or A is different in the three complexes, but the rules
of the $(1,3_1)$ diagram of Fig.~\ref{fig2}(b) for hooking up the
states are such that one necessarily ends up with the six bands
forming a connected group in the sense of Ref.~\cite{michel-prb99}.

One noteworthy difference with respect to the spinless case of
Fig.~\ref{fig9}(a) is that it is not possible to eliminate all the
Weyl nodes between $\Gamma$ and A.  The minimum number of such
crossings per 6-band complex is two, and it requires that at both TRIM
the triple node lies in energy between the two single nodes; the two
unremovable unpinned crossings then occur between the red and green
branches. This scenario is almost realized in the upper complex shown
in Fig.~\ref{fig9}(b), except for an additional accidental crossing
between the two lowest bands. Crossings that ``can be moved but not
removed'' while preserving the symmetry of the Hamiltonian were first
discussed in Refs.~\cite{michel-epl00,zak-jpa02}.

\section{\label{app:splitdir} Effect of twofold rotational symmetry on
  the Zeeman splitting of triple Weyl nodes in tellurium }

The point group of trigonal Te contains a pure 2-fold rotation
operation~\cite{asendorf-jcp57,nussbaum-procIRE62} that leaves
invariant the points $\Gamma$ and A where triple Weyl nodes occur (the
invariant lines are $\Gamma$K and AH).  That $2_0$ symmetry was not
taken into account when analyzing in Sec.~\ref{sec:trbreak} the
splitting of a triple node by a Zeeman field.  In this appendix, we
show that its presence pins the three single nodes that split from the
triple node to the $\Gamma$MLA planes in the BZ.

To proceed we need to find the constraints imposed by the $2_0$
symmetry on the unperturbed effective Hamiltonian of a triple node.
The first step is to write down the matrix $C_{2_0}$ representing the
$2_0$ operation in the basis of the two states forming the triple
node.  Since that operation sends $\kz$ into $-\kz$ and the function
$g(0,0,q_z)$ multiplying $\sigma_z$ in $H_{\rm eff}(0,0,\kz+q_z)$ is
odd [see \eq{g-00qz}], it follows that the basis states transform one
into the other with some phase factors,
\begin{equation}
C_{2_0}=\left(\begin{array}{cc}
0&e^{i\phi_2}\\
e^{i\phi_1}&0
\end{array}  \right).
\end{equation}
From $\left(2_0\right)^2=(-1)^F$, we obtain for spinful electrons the
constrain $e^{i(\phi_1+\phi_2)}=-1$, while $\phi_1-\phi_2$ depends on
the arbitrary choice of phases for the two basis states.  Since we
have not fixed those phases anywhere so far we are free to choose
$\phi_1=\phi_2=-\pi/2$, leading to
\begin{equation}
C_{2_0}=-i\sigma_x.\label{eq:C2rot}
\end{equation}

The $2_0$ operation transforms the wavevector measured relative to the
nodal point $\Gamma$ or A as $q_+\leftrightarrow q_-$ and
$q_z\rightarrow-q_z$, and using \eq{rot-symm} we find as the
invariance condition for the unperturbed effective Hamiltonian
$C_{2_0} H_{\rm eff}(q_+,q_-,q_z)C_{2_0}^{-1}=H_{\rm
  eff}(q_-,q_+,-q_z)$.
Inserting \eq{ham-eff} for $H_{\rm eff}$ in this relation and using
\eq{C2rot} leads to the constraints
\begin{subequations}
\begin{eqnarray}
f(q_-,q_+,-q_z)&=&f^*(q_+,q_-,q_z)\\
g(q_-,q_+,-q_z)&=&-g(q_+,q_-,q_z)
\end{eqnarray}
\end{subequations}
which imply, for the expansion coefficients in \eq{expansion},
\begin{subequations}
\begin{eqnarray}
A_{n_1n_2n_3}&=&(-1)^{n_3}A_{n_1n_2n_3}^* \\
B_{m_1m_2m_3}&=&(-1)^{m_3+1}B_{m_2m_1m_3}. 
\end{eqnarray}
\end{subequations}
It follows from the first condition that $A_{n_1n_2n_3}$ is real
(purely imaginary) when $n_3$ is even (odd), and from the second
combined with \eq{B-cond} that $B_{m_1m_2m_3}$ is real (purely
imaginary) when $m_3$ is odd (even).

When applied to the unperturbed effective Hamiltonian of a triple Weyl
node in trigonal Te [\eq{ham_wp3} with $M_z=0$], the above constraints
from $2_0$ symmetry imply that the parameters $\alpha=B_{001}$,
$a=A_{300}$ and $b=A_{030}$ are real, while $c=B_{300}$ and
$\beta=A_{001}$ are purely imaginary.  As a result the previously
complex quantities $\wt a$ and $\wt b$ appearing in \eq{phiq} have now
become real, and $\wt\gamma$ has become purely imaginary. The latter
follows from the fact that $2_0$ symmetry renders $\gamma$ purely
imaginary:
\bea
\gamma&=&\me{u}{\tau_z}{v}=\me{C_{2_0}u}{ C_{2_0} \tau_z C_{2_0}^{-1}}{C_{2_0}v}\nn
&=&-\me{v}{\tau_z}{u}=-\gamma^*,
\eea
where $C_{2_0}\ket{v}=-i\ket{u}$ and $C_{2_0}\ket{u}=-i\ket{v}$
according to \eq{C2rot}, and $ C_{2_0} \tau_z C_{2_0}^{-1}=-\tau_z$
according to the algebra for spin-$\frac{1}{2}$
rotations~\cite{sakurai-book94}.  Under these circumstances the real
part of \eq{phiq} reduces to $(\wt a+\wt b)\cos(3\phi)=0$, which for
$\wt a\neq -\wt b$ has six inequivalent roots
\begin{equation}
\phi=\frac{\pi}{6}+l\frac{\pi}{3}, \quad l=0,1,\ldots,5,
\end{equation}
and from the imaginary part of \eq{phiq} we find, using
$\sin(3\phi)=(-1)^l$,
\begin{equation}
q_\perp^3=(-1)^{l}\frac{ M_z\mathrm{Im}[\wt \gamma]}{\wt b-\wt a}.
\end{equation}
Depending on the various parameters, the three physical solutions with
a positive $q_\perp$ are the ones with either even or odd values
of~$l$. These two possibilities realize the two types of 3-fold
symmetric patterns where the split Weyl nodes lie on the $\Gamma$MLA
planes.

\end{document}